\documentclass[10pt,pre,aps,showkeys,showpacs,twocolumn,amsmath,amssymb,superscriptaddress,nofootinbib]{revtex4-1}
\usepackage{mathtools}
\usepackage{bm}
\usepackage{mathrsfs}
\usepackage[normalem]{ulem}
\usepackage{color}
\usepackage{kvsetkeys}
\usepackage{graphicx}

\begin{document}

\title{The structure of deterministic mass and surface fractals:\\
theory and methods of analyzing small-angle scattering data
}

\author{A. Yu. Cherny}
\email[Corresponding author, e-mail:~]{cherny@theor.jinr.ru}
\affiliation{Joint Institute for Nuclear Research, Dubna 141980, Russian Federation}

\author{E. M. Anitas}
\affiliation{Joint Institute for Nuclear Research, Dubna 141980, Russian Federation}
\affiliation{Horia Hulubei, National Institute of Physics and Nuclear Engineering, RO-077125 Bucharest-Magurele, Romania}

\author{V. A. Osipov}
\affiliation{Joint Institute for Nuclear Research, Dubna 141980, Russian Federation}

\author{A. I. Kuklin}
\affiliation{Joint Institute for Nuclear Research, Dubna 141980, Russian Federation}
\affiliation{Laboratory for Advanced Studies of Membrane Proteins, Moscow Institute of Physics and
Technology, Dolgoprudniy 141700, Russian Federation}

\date{\today}

\begin{abstract}
Small-angle scattering (SAS) of X-rays, neutrons or light from ensembles of randomly oriented and placed deterministic fractal structures are studied theoretically. In the standard analysis, a very few parameters can be determined from SAS data: the fractal dimension, and the lower and upper limits of the fractal range. The self-similarity of deterministic structures allows one to obtain additional characteristics of their spatial structures. The paper considers models which can describe accurately SAS from such structures. The developed models of deterministic fractals offer many advantages in describing fractal systems, including the possibility to extract additional structural information, an analytic description of SAS intensity, and effective computational algorithms. Generalized Cantor fractals and few of its variants are used as basic examples to illustrate the above concepts and to model physical samples with mass, surface, and multi-fractal structures. The differences between the deterministic and random fractal structures in analyzing SAS data are emphasized. Several limitations are identified in order to motivate future investigations of deterministic fractal structures.
\end{abstract}


\maketitle

\section{Introduction}

\hyphenation{USANS}

Recent advances in nanotechnology and materials science allows researchers and engineers to fabricate structures at nano and micro-scales with predefined functions and properties
\cite{Zhao2018InvestigationMuscle,Sha2018FacileCells,Acosta2018CharacterizingComposites,Sedighi2018FabricationNanoparticles,Bica2018EffectMembranes,Bica2018MagneticIron,Zhang2018PreparationSurface}.
Small-angle scattering (SAS) of X-rays or neutrons (SAXS, SANS) is a widely used non-destructive technique~\cite{Feigin1987StructureScattering,Lindner2002NeutronsMatter,Gille2013ParticleApplications} for studying their structural properties in the range from 1~nm to about 1~$\mu$m (ultra SANS). It yields information on size and shape of particles for a broad range of materials, such as polymers, metallic and organic systems or liquid crystals. The contrast variation method \cite{stuhrmann70} in SANS is of particular importance in structural biology, since it allows us to emphasize or conceal certain features of complicated biological objects~\cite{Svergun1999RestoringAnnealing,Svergun2003Small-angleSolution}. In addition, a combination of SANS/SAXS with high-resolution data from protein crystallography and nuclear magnetic resonance measurements can provide a complete image of complex biological systems~\cite{Skou2014Synchrotron-basedSolution,Madl2011NMRSolution}.

Physically, SAXS occurs due to interactions of the incident radiation with electrons in the atoms of the involved sample, while for SANS, the scattering arises from the interactions between neutrons and atomic nuclei.
Since the wavelengths of X-rays and thermal neutrons, used in a typical experiment, are of the same order of magnitude, the data analysis for SAXS and SANS can be interchanged, and generally, the same theoretical models are applicable to the both methods.

Experimentally, the SAS technique
produces the differential elastic cross section per unit solid angle~\cite{Feigin1987StructureScattering}, that is, the scattering intensity $I(q) \equiv (1/V') \mathrm{d}\sigma / \mathrm{d}\Omega$ as a function of the momentum transfer $q = 4 \pi \lambda^{-1} \sin \theta$, where $V'$ is the irradiated volume, $\lambda$ is the radiation wavelength and $2\theta$ is the scattering angle. The scattering intensity is controlled by the spatial density-density correlations in the sample under study, and it is the starting point of structural analysis, which, in principle, enables us to find the pair correlation function
by an indirect Fourier transform~\cite{Glatter1977AData,Brunner-Popela1997Small-AngleTechnique,Bergmann2000SolvingBSSA}.

Fractal geometry~\cite{Mandelbrot1982TheNature,Gouyet1996PhysicsStructures} has been proved to be a very useful ``language'' for describing the scaling behavior of objects that show a kind of self-similarity, either an exact (deterministic) one, where an intrinsic pattern repeats itself exactly under a transformation of scale, or a statistical (random) one, where the statistical properties are unchanged under scaling. Most of the fractal structures, found in nature or composed, are self-similar only statistically, see, e.g., recent studies of growth of bacterial clusters~\cite{Das2017ExtracellularFrom}, water resistant cellulose -- TiO$_{2}$ composites \cite{Garusinghe2018WaterPhotocatalysis}, aqueous graphene oxide with molecular surfactants and polymers \cite{McCoy2018BulkStability} and  resorcinol formaldehyde aerogels \cite{Alshrah2018NanostructureTechnique}.

In recent years, various deterministic nano and micro structures have been fabricated artificially, such as two-dimensional Cantor sets~\cite{Cerofolini2008FractalNanotechnology}, molecular hexagonal Sierpinski gaskets~\cite{Newkome2006NanoassemblyGasketquot}, three-dimensional Menger sponges~\cite{Mayama2006MengerMethod}, octahedral structures~\cite{Berenschot2013FabricationSilicon}, or Sierpinski triangles ~\cite{Li2017ConstructionOrder,Li2015Sierpinski-triangleGroup,Zhang2018SierpinskiAu111}. One of the main incentives for the synthesis of deterministic structures is to create and investigate materials with unique physical properties. To this end, various physical properties of deterministic fractals, such as propagation of electromagnetic waves~\cite{Tarasov2015ElectromagneticFractals}, radii of gyration~\cite{Dolgushev2016ExtendedApplications}, radiative heat transfer \cite{Nikbakht2017RadiativeStructures}, and NMR relaxation~\cite{Markelov2018NMRFractals}, are studied.

Most of the deterministic structures mentioned above are composed of small number of micro-fractal objects, while samples suitable for SAS experiments are quite difficult to create, since they must contain a \emph{macroscopic} number of the fractal objects. Recently there has been progress in developing new devices and methods to overcome this issue, such as high resolution 3$d$ and 4$d$ printing and various lithographic techniques~\cite{Wang20173DProspective,Momeni2017APrinting,Choong20174DStereolithography,Lafuente20183DAgents,Ngo2018AdditiveChallenges,Na2018EffectBioink,Choi2018EffectProliferation,Koo20183DRegeneration}.

One of the advantages of SAS technique is its ability to distinguish, on the one hand, between mass~\cite{Teixeira1988Small-angleSystems} and surface fractals~\cite{Bale1984Small-AngleProperties}, and, on the other hand, between random and deterministic fractals~\cite{Schmidt1991Small-angleSystems}. In the first case, the difference is accounted for by the value of the scattering exponent $\tau$ of the SAS curve in the fractal region
\begin{equation}
I(q) \sim q^{-\tau},
    \label{eq:int}
\end{equation}
with
\begin{equation}
\tau=\begin{dcases}
D_{\mathrm{m}},       &\text{for mass fractals},\\
2d - D_{\mathrm{s}},  &\text{for surface fractals},
\end{dcases}
    \label{eq:tau}
\end{equation}
where $d$ is the Euclidean dimension of the space in which the fractal is embedded, and $D_{\mathrm{m}}$ and $D_{\mathrm{s}}$ are the correspondent fractal dimensions. Recall that in a two-phase configuration in a $d$-dimensional space, there are a phase (``mass'') with dimension $D_{\mathrm{m}}$ and its complement (``pores''), which is another phase with dimension $D_{\mathrm{p}}$. Then the boundary between the two phases forms also a set (``surface'') with dimension $D_{\mathrm{s}}$. By definition, for a mass fractal $D_{\mathrm{s}} = D_{\mathrm{m}}<d$ and $D_{\mathrm{p}} = d$, while for a surface fractal, we have $D_{m} = D_{p} = d$ and $d-1 < D_{\mathrm{s}} < d$. Experimentally, this is interpreted as follows: if the measured value of $\tau$ is such that $\tau < d$ then the object is classified as a mass fractal with dimension $D_{\mathrm{m}} = \tau$, while if $d< \tau < d + 1$ the object is a surface fractal with $D_{\mathrm{s}} = 2d-\tau$.

For random fractals, the scattering intensity is characterized by a simple power-law decay (\ref{eq:int})~\cite{Martin1987ScatteringFractals,Schmidt1991Small-angleSystems}, while for deterministic fractals, the intensity consists of a superposition of maxima and minima on a simple power-law decay~\cite{Schmidt1986CalculationFractals,Cherny2010ScatteringFractals,Cherny2011DeterministicData,Anitas2015MicroscaleFractals,Cherny2017ScatteringFractals,Cherny2011DeterministicData}. This behavior is known as the generalized power-law decay (GPLD). However, wavelength spread of the beam, limits of the collimation, detector resolution, and the presence of polydispersity lead to smearing of the intensity curves, and the maxima and minima could be so smoothed that the GPLD becomes a simple power-law decay (\ref{eq:int}). In this case, one cannot distinguish between the scattering intensity from a strong polydisperse system of deterministic fractals and a system of simple random fractals of the same dimension.

We make some historical remarks. The simple power-law decay (\ref{eq:int}) was explained in early eighties. Technically, the scattering intensity is the Fourier transform of the pair distribution function (see details in Sec.~\ref{sec:fr_dim} below), which obeys the power-law dependence in \emph{real} space $g(r)\sim r^{D_{\mathrm{m}}-d}$ for a \emph{mass} fractal of dimension $D_{\mathrm{m}}$~\cite{forrest79,witten81}. This leads to the power-law (\ref{eq:int}) for the scattering intensity in \emph{momentum} space with the exponent $\tau=D_{\mathrm{m}}$~\cite{Sinha1984}. Bale and Schmidt \cite{Bale1984Small-AngleProperties} were first to derive the power-law exponent $\tau=2d-D_{\mathrm{s}}$ for \emph{surface} fractals as a generalization of the Porod law \cite{Porod1951} that dictates the value of exponent at large momentum to be $\tau=d+1$ for a smooth boundary between two phases with ``usual'' dimension $D_{\mathrm{s}}=d-1$. The pair distribution function for random surface fractals was studied recently in Ref.~\cite{Besselink16}. This results were successfully applied to experimental study of \emph{random} fractals. However, the obtained parameters were limited by the fractal dimension, determined from the intensity slope on double logarithmic scale, and the size of the fractal range, determined from the boundaries of the fractal region.

Models of SAS from \emph{deterministic} fractals were considered first by Schmidt \textit{et al.}~\cite{Schmidt1986CalculationFractals} in the middle eighties, but the true potential of the models was recognized much later, starting from 2010 when Cherny \textit{et al.} derived analytic expressions of SAS from deterministic mass \cite{Cherny2010ScatteringFractals,Cherny2011DeterministicData} and surface \cite{Cherny2017ScatteringFractals,Cherny2017Small-angleSnowflake} fractals. In Ref.~\cite{Cherny2010ScatteringFractals}, a model of generalized Cantor fractal with a controlled fractal dimension of $0<D_{\mathrm{m}}<3$ was proposed, and in Ref.~\cite{Cherny2011DeterministicData} a generalized Viksek model, for which the scattering amplitudes are calculated analytically. These models are mass deterministic fractals with one scale factor, the value of which determines the dimension of the fractal. As shown in Refs.~\cite{Cherny2017ScatteringFractals,Cherny2017Small-angleSnowflake}, surface fractals can be represented as sums of finite iterations of mass fractals of the same dimension, and a model of generalized Cantor surface fractal with the surface dimension $2<D_{\mathrm{s}}<3$ is proposed.

These newly proposed exactly solvable models allows one to calculate the model scattering intensities and not only verify the known properties of small-angle scattering on surface and mass fractals but also observe new previously unknown patterns of scattering, which are inherent in deterministic surface and mass fractals. In particular, the papers suggested clear recipes on how to \emph{extract additional structural parameters} from experimental SAS data, such as the fractal dimension, iteration number, scaling factor, the number and size of the basic units forming the fractal as well as the overall size of the fractal. The proposed models allows a deeper understanding of many fractal properties; in particular, it is shown that surface fractals can be represented as \emph{sums of finite mass-fractal iterations of the same dimension}.
Thus, the main geometrical parameters of a large class of deterministic fractals can be determined from SAS experiments with the help of deterministic models.

In this mini-review, we present and analyze the main deterministic models existing in the literature with a focus on how to extract structural information from SAS experimental data. First, we give a brief introduction to the theory of fractals and the SAS technique. Then, we simulate numerically SAS from the two-dimensional generalized mass Cantor fractal (GMCF). This is the starting point for investigations more complex models belonging to other classes of fractals, such as surface, fat, and multi-fractals. Finally, the algorithms for extracting structural informations are reviewed and discussed in details, including key advantages and limitations of the presented approach.

\section{Theoretical background}
\subsection{Fractal dimensions}
\label{sec:fr_dim}

One of the most important characteristic of fractal is the fractal (Hausdorff) dimension~\cite{Hausdorff1918DimensionMa}. Roughly speaking, it is a measure of how much space a set occupies near each of its points. Mathematically, the Hausdorff dimension, defined for arbitrary set, is introduced by an abstract definition (see below) based on measures \cite{Hausdorff1918DimensionMa,Gouyet1996PhysicsStructures,Falconer2003FractalApplications}. Let us consider $A$ a subset of $n$-dimensional Euclidean space and $\{V_{i}\}$ a covering of $A$ such that $|V_{i}| \leqslant a$, where $|V|$ denotes the diameter of the set $V$, which is defined by $|V| = \mathrm{sup} \{ |x-y|: x, y \in V \}$. Then by definition, the $\alpha$-dimensional Hausdorff measure $\mathrm{m}^{\alpha}(A)$ of $A$ is given by
\begin{equation}
\mathrm{m}^{\alpha}(A) \equiv \lim_{a \rightarrow 0} \inf_{\{V_{i}\}}\sum_{i}a_{i}^{\alpha},\quad\alpha > 0,
    \label{eq:malpha}
\end{equation}
where the infimum is taken over all possible coverings. In general, $\mathrm{m}^{\alpha}(A)$ is infinite and $\alpha$ is not integer. Thus, the Hausdorff dimension $D$ of the set $A$ is
\begin{equation}
D \equiv \inf\{ \alpha: \mathrm{m}^{\alpha}(A) = 0 \} = \sup\{\alpha: \mathrm{m}^{\alpha}(A) = + \infty \},
    \label{eq:hd}
\end{equation}
that is, $D$ is the value of $\alpha$ for which the Hausdorff measure jumps from zero to infinity.

The rigorous mathematical definition (\ref{eq:hd}) is not very convenient in practice. We give a more convenient heuristic definition, accepted in the literature. If $N$ is the ``minimal" number of balls of radius $a$ that cover the fractal of length $L$  then the fractal dimension is given by the asymptotics
\begin{equation}
N \sim \left( L/a \right)^{D}
    \label{eq:N}
\end{equation}
in the limit $a\to 0$. On the other hand, if $a=\mathrm{const}$ then $N \sim L^{D}$. This suggests the mathematically rigorous
``mass-radius'' approach. Consider the mass $M(r)$, i.e., the total fractal measure such as mass, surface or volume, embedded within a ball of dimension $d$ and radius $r$ centered on a point that belongs to the fractal. We have
\begin{equation}
M(r) = A(r) r^{D}
    \label{eq:Mr}
\end{equation}
 with $\ln A(r) / \ln r \rightarrow 0$ when $r \rightarrow \infty$. In most of the cases, the ``mass-radius'' definition of the fractal dimension is equivalent to the above definition (\ref{eq:hd}), see the details in Ref.~\cite{Gouyet1996PhysicsStructures}.

Quite often deterministic fractals are generated by iterative rules. This assumes the existence of an initial set (initiator) as well as an iterative operation (generator). The fractal iteration is the number of iterative operations. For instance, one can construct a single-scale deterministic fractal with an initiator of ``usual" form like ball or cube of size $L$ and a generator consisting of $k$ scaled down copies of the initiator with the scaling factor $\beta{}$. Then we can write for the total fractal mass
\begin{equation}
M(L)=k M(\beta{}L).
    \label{eq:ML}
\end{equation}
The ``mass-radius'' relation \eqref{eq:Mr} yields
\begin{equation}
k \beta{}^{D} = 1.
    \label{eq:Mrv2}
\end{equation}
 This is a fundamental relation for obtaining the fractal dimension of arbitrarily single-scale deterministic fractal. A multiple-scale fractal (multifractal) is generated with many scaling factors $\beta_{i}$ and $k_{i}$ copies at each iteration. Equation \eqref{eq:Mrv2} can be generalized to
\begin{equation}
\sum_{i}k_{i}\beta_{i}^{D} = 1.
    \label{eq:Mrv3}
\end{equation}

Experimentally, any fractal structure can be realized within a finite range of scales. For this reason, the mass-radius relation (\ref{eq:Mr}) is fulfilled within a range $l_{\mathrm{min}} \lesssim r \lesssim L$, which is called \emph{fractal range}. This implies that in the reciprocal momentum space, the observable fractal region lies within
\begin{equation}
2\pi/L \lesssim q \lesssim  2\pi/l_{\mathrm{min}},
    \label{eq:q}
\end{equation}
for which Eq.~\eqref{eq:int} holds. This is because the SAS intensity is proportional to the Fourier transform of the \emph{pair distribution function} $g(\bm{r})$ (see below Sec~\ref{sec:SASmeth}), which describes the spatial correlations between particles inside the fractal and gives the probability density to find a particle at distance $\bm{r}$ apart from another particle, provided the position of the latter one is known. Then, for a radially symmetric distribution function, the total number of particles inside a sphere of radius $r$ is $N(r) \sim \int_{0}^{r}g(r') {r'}^{d-1}\mathrm{d}r'$. For a mass fractal of dimension $D_{\mathrm{m}}$, we have $N(r) \sim r^{D_{\mathrm{m}}}$ from Eq.~\eqref{eq:Mr} and thus find
\begin{equation}
g(r) \sim r^{D_{\mathrm{m}}-d},\quad \mathrm{for}\ l_{\mathrm{min}}\lesssim r \lesssim L.
    \label{eq:gr}
\end{equation}
By using Erd\'ely's theorem \cite{erdelyi56:book} for asymptotic expansions of Fourier integrals, Eq.~\eqref{eq:int} is recovered.

\subsection{SAS method}
\label{sec:SASmeth}

In this section, we mainly follow our work \cite{Cherny2011DeterministicData}.

Let us consider an incident beam of neutrons or X-rays hiting a sample. The total irradiated volume of the sample $V'$ contains a macroscopic number of scattering microscopic units with the scattering length $b_{j}$. Then the scattering amplitude $A(\bm{q})$ takes the form \cite{Feigin1987StructureScattering} $A(\bm{q}) = \int_{V'}\rho_{s}(\bm{r}) e^{-i \bm{q} \cdot \bm{r}}\mathrm{d}\bm{r}$, where $\rho_{s}(\bm{r}) = \sum_{j}b_{j}\delta (\bm{r}-\bm{r}_{j})$ is the scattering length density (SLD), $\bm{r}_{j}$ are the microscopic-unit positions and $\delta$ is the Dirac $\delta$-function. The differential cross section of the sample can be written through the amplitude as $\mathrm{d} \sigma/\mathrm{d}\Omega = |A(\bm{q})|^{2}$.

We assume that the sample is a two-phase system consisting of homogeneous units of ``mass'' density $\rho_{m}$ immersed into a solid matrix of ``pore'' density $\rho_{p}$. A constant density shift across the sample plays a role only at small values of $q$, which are usually beyond  the instrumental resolution. Then by subtracting the ``pore'' density, we obtain a system where the units of the density $\Delta \rho = \rho_{m} - \rho_{p}$ are ``frozen'' in a vacuum. The density $\Delta \rho$ is called the scattering contrast. If the sample consists of many identical rigid objects of volume $V$ with concentration $n$ and whose spatial positions and orientations are uncorrelated, the scattering intensity can be written as
\begin{equation}
I(q) \equiv \frac{1}{V'}\frac{\mathrm{d} \sigma}{\mathrm{d}\Omega}=n |\Delta \rho|^{2}V^{2}\left\langle |F(\bm{q})|^{2} \right\rangle,
    \label{eq:intv2}
\end{equation}
where the normalized scattering amplitude (formfactor) of the object is introduced
\begin{equation}
F(\bm{q}) \equiv \frac{1}{V}\int_{V}e^{-i \bm{q} \cdot \bm{r}}\mathrm{d}\bm{r}.
    \label{eq:Fq}
\end{equation}
It satisfies the normalization condition $F(0) = 1$. It follows from Eq.~\eqref{eq:intv2} that
\begin{equation}
I(0) = n |\Delta \rho|^{2}V^{2}.
    \label{eq:int0}
\end{equation}
The object volume can be determined experimentally from this useful relation provided their concentration $n$ and the contrast $\Delta \rho$ are known.

The above notation $\left\langle \cdots \right\rangle$ represents the ensemble averaging over all orientations of the object. If the probability of any orientation is the same, it can be calculated in three dimensions ($d=3$) by integrating over the solid angle in the spherical coordinates $q_{x} = q \cos \phi \sin \theta$, $q_{y} = q \sin \phi \sin \theta$ and $q_{z} = q \cos \theta$:
\begin{equation}
\left\langle  f(q_{x}, q_{y}, q_{z}) \right\rangle = \frac{1}{4\pi} \int_{0}^{\pi} \mathrm{d}\theta \sin \theta \int_{0}^{2\pi}\mathrm{d}\, \phi f(q, \theta, \phi),
    \label{eq:averaging3d}
\end{equation}
where $f$ is an arbitrarily function. In two-dimensional space ($d=2$), the mean value takes the form
\begin{equation}
\left\langle f(q_{x}, q_{y}) \right\rangle = \frac{1}{2\pi} \int_{0}^{2\pi}f(q,\phi)\,\mathrm{d}\phi,
    \label{eq:averaging2d}
\end{equation}
where $q_{x} = q \cos \phi$ and $q_{y} = q \sin \phi$, and the volume in Eqs.~\eqref{eq:Fq} and \eqref{eq:int0} should be replaced by the object area.

It follows from the definition \eqref{eq:Fq} that the formfactor obeys the properties, quite useful when calculating the SAS curves:
\begin{itemize}
    \item $F(\bm{q}) \rightarrow F(\beta{}\bm{q})$ when the length of the particle is scaled as $L \rightarrow \beta{}L$,
    \item $F(\bm{q}) \rightarrow F(\bm{q})e^{-i \bm{q} \cdot \bm{a}}$ when the particle is translated $\bm{r} \rightarrow \bm{r} + \bm{a}$,
    \item $F(\bm{q}) = \left[ V_{I}F_{I}(\bm{q}) + V_{II}F_{II}(\bm{q}) \right] / \left( V_{I} + V_{II} \right)$, when the particle consists of two non-overlapping subsets $I$ and $II$.
\end{itemize}

If the object is a fractal consisting of
\begin{equation}
N_{m} = k^{m}
    \label{eq:Nm}
\end{equation}
``primary'' units of the same shape and size ($m$ is the iteration number), its formfactor can be written as \cite{Cherny2011DeterministicData}
\begin{equation}
F(\bm{q}) = \rho_{\bm{q}}F_{0}(\bm{q})/N_{m},
    \label{eq:fractalff}
\end{equation}
where $F_{0}(\bm{q})$ is the formfactor of the initiator (ball, cube, disk etc.), and $\rho_{\bm{q}} = \sum_{j}e^{-i \bm{q}\cdot \bm{r_{j}}}$ is the Fourier component of the density of the scattering unit centers with $\bm{r}_{j}$ being their positions.

\begin{figure}
	\begin{center}
		\includegraphics[width=\columnwidth,clip]{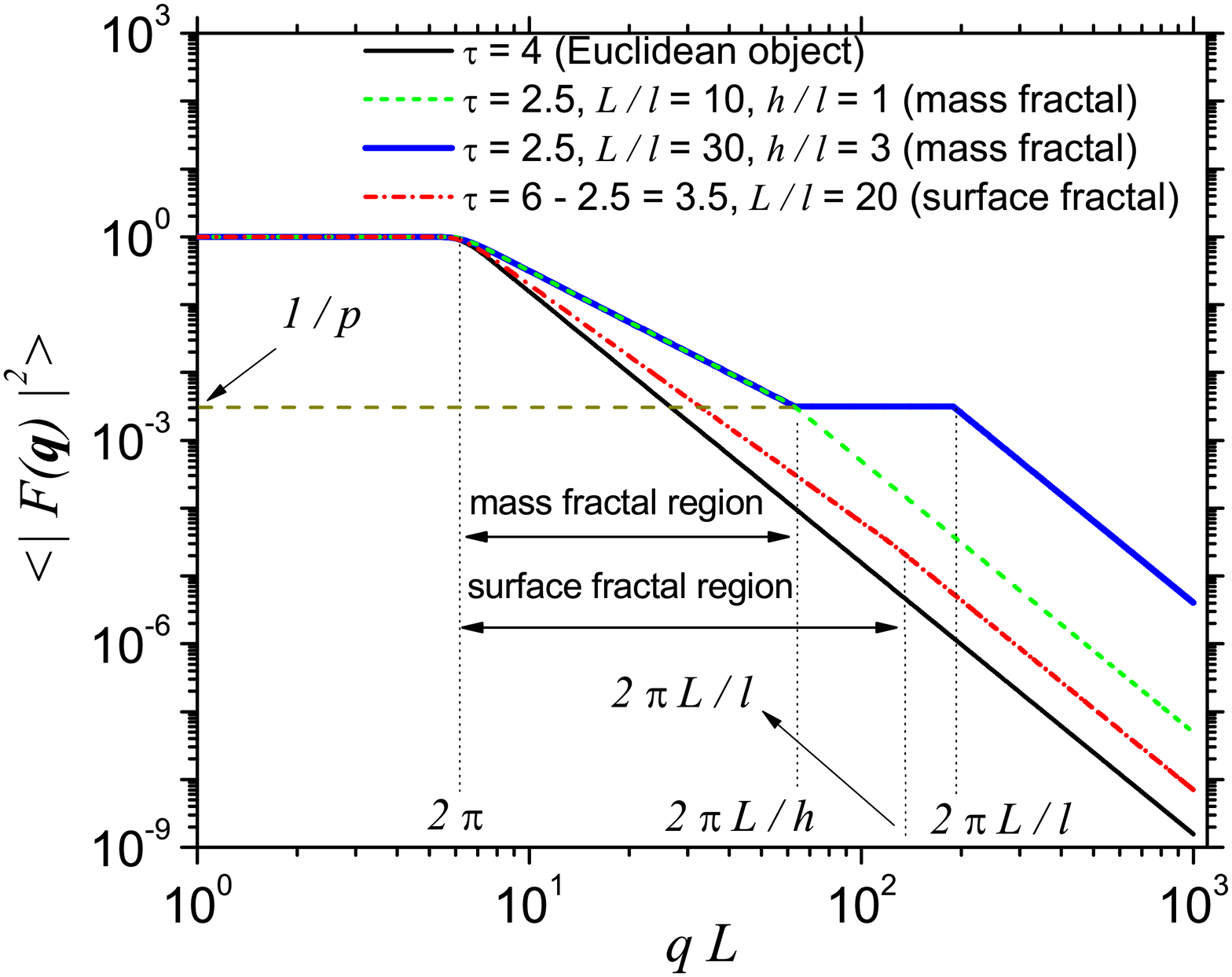}
		\caption{\label{fig:fig1} Schematic SAS intensities from regular and fractal objects in three-dimensional Euclidean space (in units of $n|\Delta \rho|^{2}V^{2}$) as a function of the momentum transfer (in units of $1/L$). For regular objects the intensity \eqref{eq:F0v2} (solid black line) is characterized by the presence of two regions: Guinier (when $q \lesssim 2\pi/L$) and Porod (when $ q\gtrsim 2\pi/L$) with $L$ being the overall size of the object. For mass fractals, the intensity is given by Eq.~\eqref{eq:Fm} (green dashed line). If the ratio  of the distances between primary units $h$ to their size $l$ is close to one, an intermediate fractal region lies between the Guinier and Porod ones. If $h / l \gg 1$, a ``shelf'' of the value $1/p$, immediately following the fractal region, appears (blue solid line). Here $p$ is of order $\left( L / h \right)^{D_{\mathrm{m}}}$. In the SAS intensity of surface fractal \eqref{eq:Fs} (red dot-Dashed line), the fractal region is immediately followed by the Porod one.}
\end{center}
\end{figure}

Generally, for a ``regular'' object of size $L$ embedded in $d$-dimensional Euclidean space, the normalized SAS intensity takes the form \cite{Porod1951}
\begin{equation}
\left\langle |F_{0}(\bm{q})|^{2} \right\rangle \simeq
\begin{dcases}
    1,&q \lesssim 2\pi / L,\\
    \left( q L / 2 \pi \right)^{-\tau},&2\pi / L \lesssim q,
\end{dcases}
    \label{eq:F0v2}
\end{equation}
with $\tau=d+1$. This is the well-known Porod law\footnote{Note that if a ``regular'' $d$-dimensional object is immersed into \emph{three-dimensional} space, the averaging \eqref{eq:averaging3d} yields $\tau = 4$, $2$, and $1$ for $d = 3$, $2$, and $1$, respectively. This allows us to find the object dimension experimentally.}. The main feature is the presence of two main structural regions: the Guinier region ($q \lesssim 2\pi / L$) and the Porod region  ($q \gtrsim 2\pi /L $), from which we can extract information about the overall size of the object.
Fig.~\ref{fig:fig1} shows the behaviour of the normalized intensity for typical two- and three-dimensional objects.

For simplicity, we choose a disk (for $d=2$) or ball (for $d=3$) as the initiator; its formfactor $F_0$ is radially symmetric.  Substituting Eq.~\eqref{eq:fractalff} into Eq.~\eqref{eq:intv2} yields
\begin{equation}
I(q) = I(0) S(q) |F_{0}(q)|^{2}/ N_{m},
    \label{eq:intsf}
\end{equation}
where $S(q)$ is the structure factor defined by
\begin{equation}
S(q)\equiv \left\langle \rho_{\bm{q}} \rho_{-\bm{q}} \right\rangle / N_{m}.
    \label{eq:sf}
\end{equation}
It is related to the pair distribution function through
\begin{equation}
S(q) = 1 + \frac{N_{m} - 1}{L^d}\int_{V} \mathrm{d}\bm{r}\,g(r)\exp(-i\bm{q}\cdot \bm{r})
    \label{eq:sfv2}
\end{equation}
and carries information about the relative positions of the scattering unit centers inside the fractal. Equation \eqref{eq:sf} can be rewritten as
\begin{equation}
S(q) = \frac{1}{N_{m}}\sum_{j, k = 1}^{N_{m}}\left\langle e^{-i \bm{q} (\bm{r}_{j}-\bm{r}_{k})} \right\rangle,
    \label{eq:sfv3}
\end{equation}
with $S(0) = N_{m}$. The long-range asymptotics [$S(q) \simeq 1$ for $q \gtrsim 2\pi/l_{\mathrm{min}}$] is governed by the diagonal terms in Eq.~\eqref{eq:sfv3}, because  the contribution of the non-diagonal terms tends to zero
due to \emph{the randomness of the phases}.

In practice, the objects in a sample almost always have different sizes. If we assume that \emph{their shapes are the same} [which means that their normalized scattering amplitude \eqref{eq:Fq} is the same] but their sizes vary, we come to a particular type of \emph{polydispersity}.  The sizes are supposed to be random but obey a statistical distribution. The distribution function $D_{\mathrm{N}}(l)$ is defined in such a way that $D_{\mathrm{N}}(l)\mathrm{d}l$ gives the probability of finding an object whose size falls within $\left( l, l + \mathrm{d}l\right)$.  As a model, one can take the log-normal distribution
\begin{equation}
D_{\mathrm{N}}(l) = \frac{1}{\sigma l (2 \pi)^{1/2}} e^{-\frac{\left( \ln (l / l_{0}) + \sigma^{2} / 2 \right)^{2}}{2 \sigma^{2}}},
    \label{eq:DN}
\end{equation}
where $\sigma= \left( \ln (1 + \sigma_{\mathrm{r}}^{2}) \right)^{2}$,  $l_{0} = \left\langle l \right\rangle_{D}$ is the mean value of the length, $\sigma_{\mathrm{r}} \equiv \left( \left\langle l^{2} \right\rangle_{D} -l_{0}^{2} \right)^{1 / 2} / l_{0}$ is the length \emph{relative} variance, and $\left\langle \cdots \right\rangle_{D} = \int_{0}^{\infty}\cdots D_{\mathrm{N}}(l)\mathrm{d}l$. Then the polydisperse scattering intensity is obtained by averaging Eq.~\eqref{eq:intv2} over the distribution function \eqref{eq:DN}
\begin{equation}
I(q) = n |\Delta \rho|^{2}\int_{0}^{\infty}\left\langle | F(\bm{q}) |^{2} \right\rangle V^{2}(l)D_{\mathrm{N}}(l)\mathrm{d}l.
    \label{eq:intpoly}
\end{equation}

If polydispersity is strongly developed ($\sigma_{\mathrm{r}}\gg 1$), the scattering intensity for a deterministic fractal becomes very close to the simple power-law (\ref{eq:int}). Thus,  the SAS from random fractals can be modeled with deterministic fractal of the same fractal dimension \cite{Cherny2010ScatteringFractals}.

Finally, we write down schematic expressions \cite{Cherny2017ScatteringFractals} for the fractal formfactors in $d$-dimensional space, which quite resemble real experimental curves\footnote{In real experimental curves, the breakpoints are smoothed.} when polydispersity or experimental resolution smears the oscillations. They are very useful for a qualitative understanding of their scattering properties. For a mass fractal of dimension $D_{\mathrm{m}}$, overall size $L$, and consisting of $p$ units of size $l$ separated by the characteristic minimal distance $h$,
the formfactor is given by
\begin{equation}
\left\langle |F^{\mathrm{(m)}}(\bm{q})|^{2} \right\rangle \simeq
\begin{dcases}
    1, &q \lesssim \frac{2\pi}{L}, \\
    \left( qL / 2\pi \right)^{-D_{\mathrm{m}}},&\frac{2\pi}{L} \lesssim q \lesssim \frac{2\pi}{h},\\
    1/p,& \frac{2\pi}{h}\lesssim q \lesssim \frac{2\pi}{l},\\
    \left( 1/p \right) \left( qL / 2\pi \right)^{-d-1},&\frac{2\pi}{l}\lesssim q,
\end{dcases}
    \label{eq:Fm}
\end{equation}
where $p=N_m$ is of order\footnote{Note that $h$ and $l$ are of the same order in the limit of large $m$, and, hence, $\left( L / h \right)^{D_{\mathrm{m}}}\sim \left( L / l \right)^{D_{\mathrm{m}}}$ in this limit.} $\left( L / h \right)^{D_{\mathrm{m}}}$.

A similar expression can be written for surface fractals of dimension $D_{\mathrm{s}}$, composed of scattering units of maximum size $r_{0}$ and minimum size $l$,
\begin{equation}
\left\langle |F^{\mathrm{(s)}}(\bm{q})|^{2} \right\rangle \simeq
\begin{dcases}
    1, &q \lesssim \frac{2\pi}{L}, \\
    \left( \frac{q\, r_{0}}{2\pi} \right)^{D_{\mathrm{s}}-2d},& \frac{2\pi}{L}\lesssim q \lesssim \frac{2\pi}{l},\\
    \left( \frac{r_{0}}{l}\right)^{D_{\mathrm{s}}-2d} \left(\frac{2\pi }{q l } \right)^{d+1},&\frac{2\pi}{l}\lesssim q.
\end{dcases}
    \label{eq:Fs}
\end{equation}
In equation \eqref{eq:Fs}, $r_{0}$ is of order of $L$ or smaller\footnote{If $r_{0}\ll L$, an additional ``transition'' region appears, see Sec.~\ref{subsec:monoformfactor} below. For simplicity, we do not discuss this region here.}. The relations \eqref{eq:Fm} and \eqref{eq:Fs} exhibit an intermediate (fractal) region between the Guinier and Porod regions. The scattering exponent is directly related to the fractal dimension in accordance with Eq.~\eqref{eq:tau}. Figure \ref{fig:fig1} shows the typical behavior of mass and surface fractal intensities. Note the appearance of a ``shelf'' at $1/p$ when $h / l \gg 1$. As we show below, it can play an important role in explaining SAS properties of surface fractals.

\section{Small-angle scattering from mass fractals}
\label{sec:SASmass}

For the sake of simplicity, we consider two-dimensional mass fractals.  Their construction is based on an iterative ``top-down'' approach, when an initial object is repeatedly replaced by smaller copies of itself according to an iterative rule.
The volumes appearing in Eqs.~\eqref{eq:intv2}, \eqref{eq:Fq} and \eqref{eq:int0} shall be replaced by the corresponding areas, and the averaging is taken with Eq.~\eqref{eq:averaging2d}.  In this section, we mainly follow the papers \cite{Cherny2010ScatteringFractals,Cherny2011DeterministicData}.

\begin{figure}
	\begin{center}
		\includegraphics[width=\columnwidth,trim={1cm 0 0 0},clip]{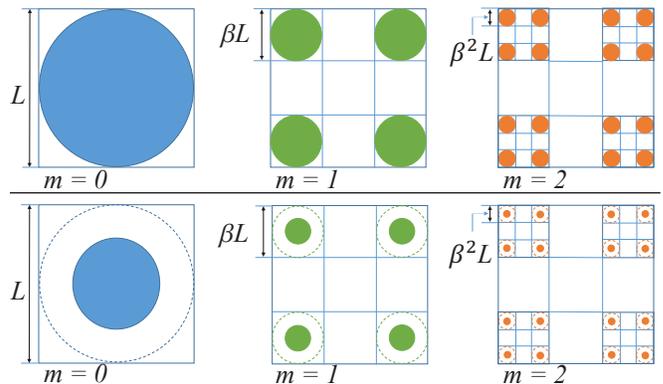}
		\caption{\label{fig:fig2} The initiator (disk) and the first two iterations
for the generalized mass Cantor fractal at scaling factor $\beta{} = 1 / 3$ and two different values of the initiator $r_0=L/2$ (upper panel) and $r_0=L/4$ (lower panel).
}
\end{center}
\end{figure}

In constructing the generalized mass Cantor fractal \cite{Cherny2010ScatteringFractals}, we start with a square of edge size $L$, in which a disk of radius $r_{0}$ (the initiator) with $0 < r_{0} \leqslant L/2$ is inscribed, such that their centers coincide (Fig.~\ref{fig:fig2}). This is the zero order iteration ($m = 0$).
We choose a Cartesian system of coordinates with the origin coinciding with the centers of the square and disk, and whose axes are parallel to the square edges. By replacing the initial disk with $k = 4$ smaller disks of radius $r_{1} = \beta{}r_{0}$, we obtain the first iteration
($m = 1$). Here $\beta{}$ is the scaling factor, obeying the condition $0 < \beta{} < 1/2$.
The positions of the four disks are chosen in such a way that their centers are given by the vectors in the plane
\begin{equation}
\bm{a}_{j} =\frac{L(1-\beta{})}{2} \{ \pm 1, \pm 1 \},
    \label{eq:a}
\end{equation}
with all combinations of the signs. The second fractal iteration ($m = 2$) is obtained by performing a similar operation on each of the $k$ disks of radius $r_{1}$. For arbitrarily iterations $m$, the total number of disks is given by Eq.~\eqref{eq:Nm} and the corresponding radii are
\begin{equation}
r_{m} = \beta{}^{m} r_0.
    \label{eq:rm}
\end{equation}
The ratio of the minimal distance between the disk centers to their diameter for each iteration is
\begin{align}\label{eq:ratio}
\frac{h}{l}=\frac{(1-\beta{})L}{2\beta{}r_0}.
\end{align}
It is important for appearing the shelf in the scattering intensity [see Eq.~(\ref{eq:Fm})].
The total area is given by
\begin{equation}
S_{m} = N_{m}\beta{}^{2m}S_{0}
    \label{eq:Am}
\end{equation}
with $S_{0} = \pi r_{0}^{2}$ being the area of the initial disk. The fractal dimension of the GCF is obtained from Eq.~\eqref{eq:N} for $m\gg1$. We find
\begin{equation}
D_{\mathrm{m}}= \lim_{m \rightarrow \infty} \frac{\ln N_{m}}{\ln (r_{0}/r_{m})} = -\frac{\ln k}{\ln \beta{}},
    \label{eq:Dm}
\end{equation}
where $N_{m}$ and $r_{m}$ are given by Eqs.~\eqref{eq:Nm} and \eqref{eq:rm}, respectively\footnote{In the rigorous mathematical consideration, we must also prove that the iterative rule results in \emph{non-empty} set when $m\to\infty$. If the initiator is a square with the edge $L$, this is obvious. However, as is emphasized in the Introduction, the physical structures are \emph{always finite} and this limit is not needed for our purposes.  What is actually needed is the fractal scaling properties within a finite range, which is realized at each \emph{finite} iteration.}. Note that the fractal dimension for the well-known two-dimensional Cantor dust is recovered for $k = 4$ and $\beta{} =1/3$.

Using the definition the formfactor \eqref{eq:Fq} and its properties listed above in Sec.~\ref{sec:SASmeth} yields
\begin{equation}
S_{1}F_{1}^{\mathrm{(m)}}(\bm{q}) = \sum_{j=1}^{k}\beta{}^{2}S_{0}F_{0}(\beta{}\bm{q})e^{-i \bm{q}\cdot \bm{a}_{j}},
    \label{eq:dfp}
\end{equation}
where the formfactor of disk is given by \cite{Svergun2003Small-angleSolution}
\begin{equation}
F_{0}(q) = 2 J_{1}(q) / q
    \label{eq:ffdisk}
\end{equation}
with $J_{1}(q)$ being the Bessel function of the first kind. By calculating the sum in Eq.~\eqref{eq:dfp} explicitly, we obtain
\begin{equation}
F_{1}^{\mathrm{(m)}}(\bm{q}) = G_{1}(\bm{q}) F_{0}(\beta{}\bm{q}),
    \label{eq:F1mf}
\end{equation}
where
\begin{align}\label{eq:G1mf}
G_{1}(\bm{q}) = \cos \frac{q_{x}L(1-\beta)}{2} \cos \frac{q_{y}L(1-\beta)}{2}
\end{align}
is called the generative function of the fractal. It depends on the relative positions of the disks inside the fractal.
The formfactor of arbitrary iteration can be found by applying repeatedly the same operation
\begin{equation}
F_{m}^{\mathrm{(m)}}(\bm{q}) = G_{1}(\bm{q})G_{1}(\bm{q}\beta) \cdots G_{1}(\bm{q}\beta{}^{m-1})F_{0}(\beta{}^{m}\bm{q}).
    \label{eq:Fmmf}
\end{equation}
Finally, the scattering intensity \eqref{eq:intv2} is obtained by calculating the mean value of $|F_{m}^{\mathrm{(m)}}(\bm{q})|^{2}$ over all direction $\bm{n}$  of the momentum transfer with Eq.~\eqref{eq:averaging2d}
\begin{equation}
I(q) / I(0) = \left\langle | F_{m}^{\mathrm{(m)}}(\bm{q}) |^{2} \right\rangle.
    \label{eq:intmf}
\end{equation}
It follows from \eqref{eq:intsf} that the renormalized structure factor $S(q)/N_m$ is given by Eqs.~(\ref{eq:Fmmf}) and (\ref{eq:intmf}) with $F_0(q)=1$. The last relation amounts to replacing the initiator by a point-like object with the $\delta$-function density distribution.

\begin{figure}
	\begin{center}
		\includegraphics[width=\columnwidth,trim={0 0 0 0},clip]{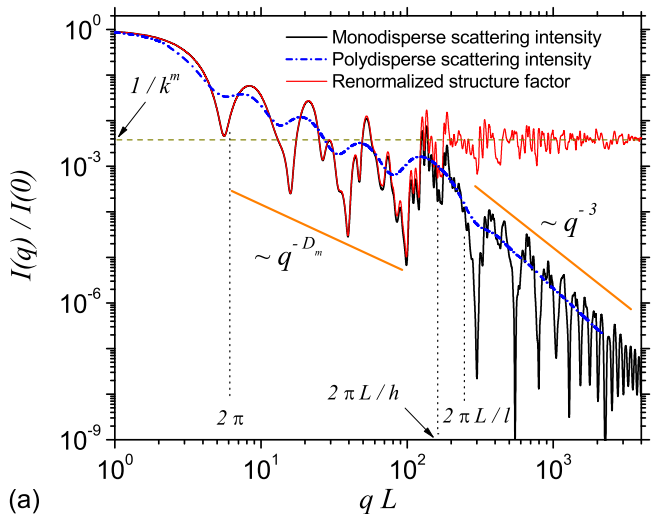}
		\includegraphics[width=\columnwidth,trim={0 0 0 0},clip]{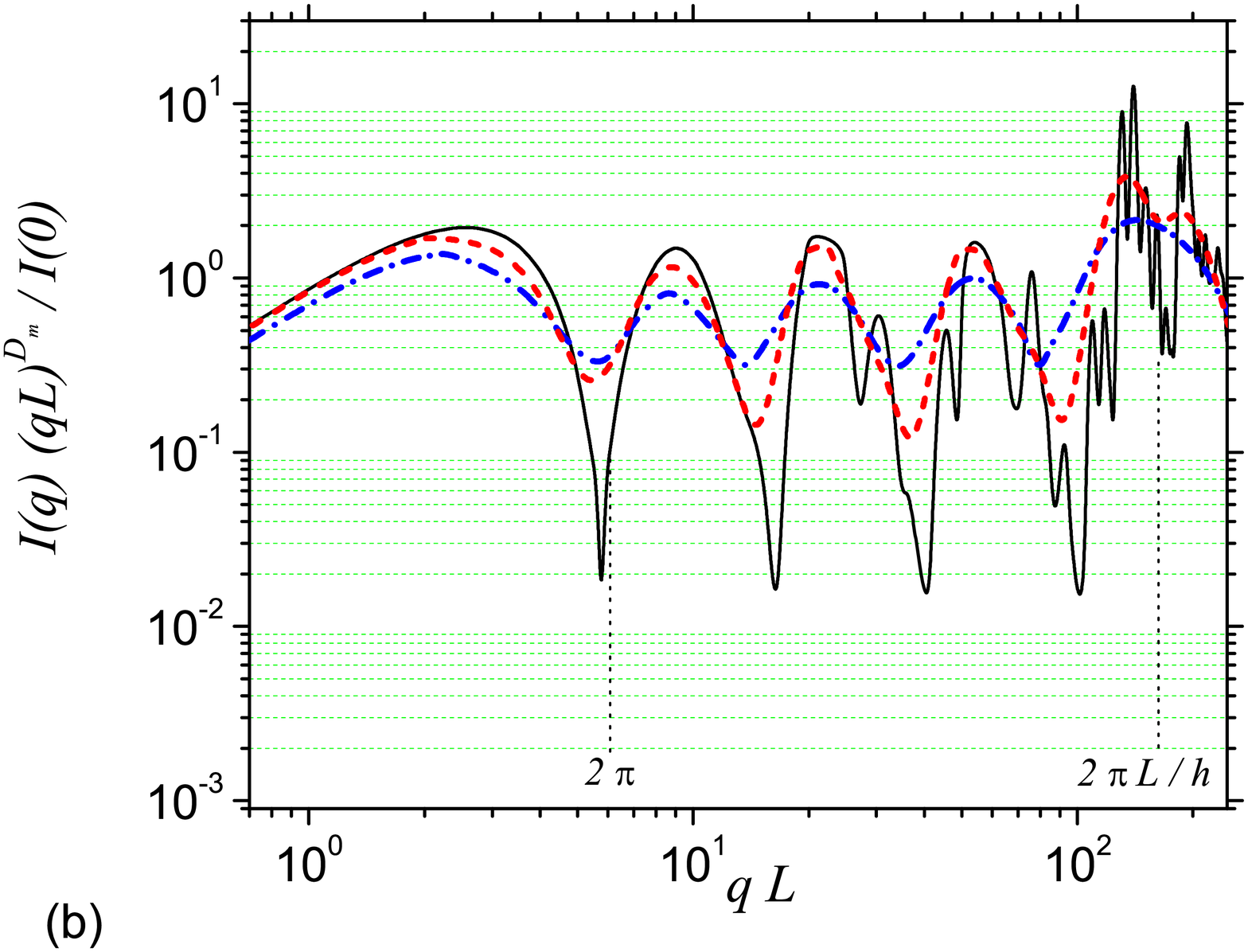}
		\caption{\label{fig:fig3} (a) Scattering intensity (\ref{eq:intmf}) (black solid line) and the renormalized structure factor $S(q)/N_m$ (red solid line) for the fourth iteration of GMCF at scaling factor $\beta{} = 0.4$, and $r_0=L/2$, for which $D_{\mathrm{m}} = 1.51\ldots$ and $h / l  = 1.5$. Blue dash-dotted line shows the intensity \eqref{eq:intpoly} in the presence of polydispersity (the log-normal distribution at the relative variance $\sigma_{\mathrm{r}} = 0.2$).
(b) The quantity $I(q) q^{D_{\mathrm{m}}}$ clearly shows the approximate periodicity in the fractal region, whose limits are indicated with the vertical black dotted lines. Red dashed and blue dash-dotted lines represent the polydisperse intensity with $\sigma_{\mathrm{r}} = 0.1$, and $\sigma_{\mathrm{r}} = 0.2$, respectively. The periodic oscillations are smeared with increasing polydispersity. }
\end{center}
\end{figure}
Figure \ref{fig:fig3}a shows the scattering intensity \eqref{eq:intmf} and the structure factor on double logarithmic scale. At $q \lesssim 2 \pi$ they are characterized by the presence of a shelf $I(q)/I(0) \sim 1$ (Guinier region).
The curvature of the intensity in Guinier region is intimately connected to the the radius of gyration: $I(q) = I(0)(1-q^{2}R_\mathrm{g}^{2}/d+\cdots)$. The explicit expression of the scattering amplitude (\ref{eq:Fmmf}) allows us to calculate it analytically for GCMF \cite{Cherny2010ScatteringFractals}, generalized Vicsek fractal \cite{Cherny2011DeterministicData}, surface \cite{Cherny2017ScatteringFractals,Cherny2017Small-angleSnowflake}, and fat fractals \cite{Anitas2014Small-angleFractals}.
When $2 \pi/L \lesssim q \lesssim 2 \pi  / l$ ($l$ is the disk diameter at given iteration), we have a fractal region of the generalized power-law decay, and whose range are determined by the maximal and minimal distances between disk centers. The exponent of the power-law decay coincide with the fractal dimension of the GMCF. The scaling factor $\beta{}$ can be obtained from the period on the logarithmic scale $\log_{10} \left( 1 / \beta{} \right)$ of the quantity $I(q)q^{D_{\mathrm{m}}}$ (Fig.~\ref{fig:fig3}b). Then the number of units $k$ in the generator can be found from Eq.~(\ref{eq:Mrv2}). Besides, the number of minima coincide with the iteration number. This allows us to obtain the total number of the fractal primary units (\ref{eq:Nm}). When $q \gtrsim 2 \pi / l$, the formfactor decays proportionally to $q^{-3}$ (Porod region) while the structure factor attains the asymptotic value $1$.

\begin{figure}
	\begin{center}
		\includegraphics[width=\columnwidth,trim={0 0 0 0},clip]{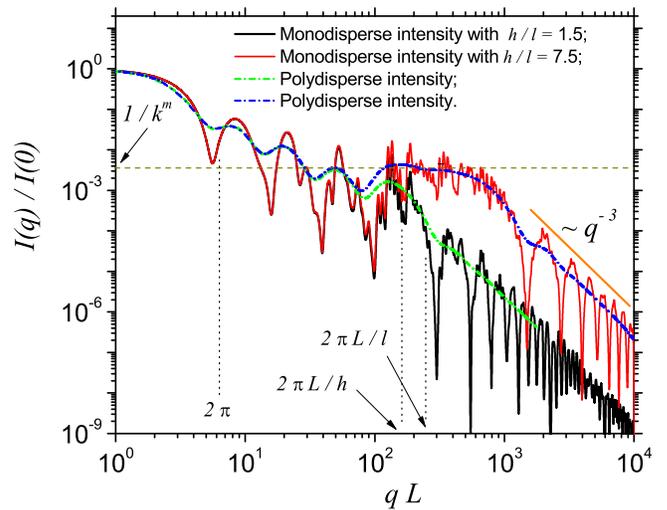}
		\caption{\label{fig:fig4} The scattering intensity for the fourth iteration of the GMCF at $\beta{} = 0.4$ ($D_{\mathrm{m}} \simeq 1.51\ldots$) for different values of the ratio $h / l$ of the minimal distances $h$ between disks' centers to their diameter. For the black line $r_{0} = L / 2$, for which $h / l = 1.5$ in accordance with Eq.~\eqref{eq:ratio}, and  for the red line $r_{0} = L / 10$ ($h / l = 7.5$). Blue and green dash-dotted lines represent the corresponding polydisperse intensities with a log-normal distribution \eqref{eq:DN} at the relative variance $\sigma_{\mathrm{r}} = 0.2$.}
\end{center}
\end{figure}

The effect of various values of the ratio $h / l$ of the distances $h$ between disks' centers to their diameter $l$
is shown in Fig.~\ref{fig:fig4}. When $r_{0} = L / 2$, $h / l = 1.5$, and the Porod region follows almost immediately the fractal one. However, when $h / l \gg 1$, the intermediate shelf appears at $1/N_m = 1/k^m$ within the range $2 \pi /h \lesssim q \lesssim 2 \pi/ l$. Thus, in analyzing experimental SAS data the beginning and end of the intermediate shelf can be used to estimate how many times the typical distances between disks is higher than their size.

\section{SAS from surface fractals}

The construction process of a surface fractal is based on the superposition of a mass fractal at various iterations \cite{Cherny2017ScatteringFractals,Cherny2017Small-angleSnowflake}. This enables us to fulfill the conditions for a surface fractal $D_{m} = D_{p} = d$ and $d-1 < D_{\mathrm{s}} < d$, discussed in the Introduction. One can make a general assumption \cite{Cherny2017ScatteringFractals} that \emph{any surface fractal can be built from a series of mass-fractal iterations of the same dimension.}

\subsection{Generalized surface Cantor fractal}
\label{subsec:monoformfactor}

\begin{figure}
	\begin{center}
		\includegraphics[width=\columnwidth,trim={1.7cm 0 6cm 0},clip]{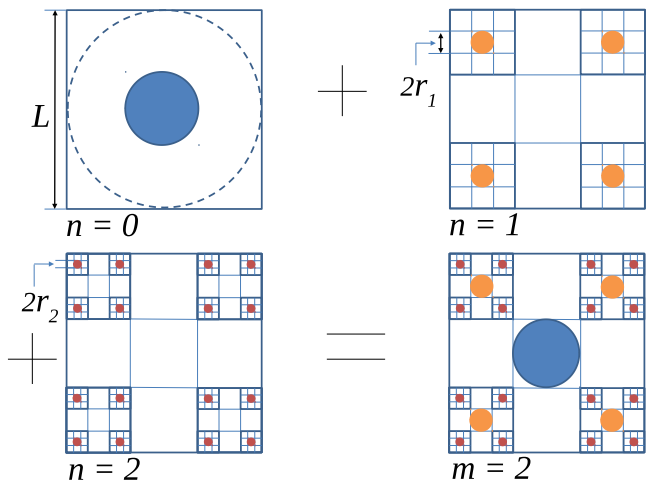}
		\caption{\label{fig:fig5} The second iteration ($m = 2$) of the two-dimensional GSCF represented as a sum of GMCF at various iterations, with disks as basic scattering units: $n = 0$ (blue), $n = 1$ (orange), and $n = 2$ (brown).}
\end{center}
\end{figure}

In this section, we compose the generalized surface Cantor fractal (GSCF) as a sum of iterations of the generalized Cantor mass fractal. Specifically,  at the $m$-th iteration the GSCF is built as a sum of GMCF of iterations from 0 to $n$, as shown in Fig.~\ref{fig:fig5}.
The GMCF is composed of disks with the same size (tending to zero at $m\to\infty$), while GSCF is composed of disks whose sizes follow a ``discrete'' power-law distribution \cite{Cherny2017ScatteringFractals}.

By the construction, the GSCF at $m$-th iteration consists of $N_{m} = 1 + k + k^{2} + \cdots k^{m}$ disks with $k = 4$ \begin{equation}
N_{m} = \left( k^{m+1} - 1 \right) / \left( k - 1 \right).
    \label{eq:NmSF}
\end{equation}
The disk of the zero iteration has  area $S_{0} = \pi r_{0}^{2}$, $k$ disks of radius $r_{1} = \beta{}r_{0}$ have area $k \beta{}^2\pi r_{0}^{2}$, $k^{2}$ disks of radius $r_{2} = \beta{}^2 r_{0}$ have area $k^{2} \beta{}^4\pi r_{0}^{2}$, etc. The total area $S_{m}$ of GSCF at $m$-th iteration is given by
\begin{equation}
S_{m} = S_{0} \frac{1 - \left( k \beta{}^{2} \right)^{m+1}}{1 - k \beta{}^{2}}.
    \label{eq:areaSF}
\end{equation}

In the limit $m \rightarrow \infty$, the total area is finite, because $k \beta{}^2 < 1$, and the Hausdorff dimension of the fractal \emph{area} of GSCF is equal to 2. Let us consider the Hausdorff dimension of the total \emph{perimeter} of the GSCF. For $m \rightarrow \infty$, the contribution of the $m$-th iteration of GMCF is given by Eq.~\eqref{eq:Dm}, since the fractal dimensions of the perimeter and area coincide for mass fractals. The contribution of the disk at $m = 0$ to the dimension of the total perimeter is equal to $1$, which imposes the lower limit for the perimeter dimension. Therefore, the Hausdorff dimension of the total perimeter of the GSCF is given by
\begin{equation}
D_{\mathrm{s}} =
\begin{dcases}
1,&0 < \beta{} < 1 / k, \\
- \ln k / \ln \beta{}, &1 / k \leqslant \beta{} < 1 /2.
    \end{dcases}
    \label{eq:Ds}
\end{equation}
The threshold value $\beta{} = 1 / k$ corresponds to $D_{\mathrm{m}} = 1$ in Eq.~\eqref{eq:Dm}, which yields $\beta{} = 1 / 4$ for $k = 4$. So, when $\beta{}$ is smaller than $1 / 4$, the total perimeter of the fractal is finite even when $m \rightarrow \infty$. The perimeter dimension \eqref{eq:Ds} satisfies the condition $1 \leqslant D_{\mathrm{s}} < 2$, as expected (see the Introduction).

The scattering amplitude of GSCF is obtained by adding the amplitudes of the mass-fractal iterations considered in Sec.~\ref{sec:SASmass}. Normalizing the result to one at $q = 0$ yields the formfactor
\begin{equation}
F_{m}^{(\mathrm{s})}(\bm{q}) = \frac{1-k\beta{}^{2}}{1-\left( k \beta{}^{2} \right)^{m+1}}\sum_{n=0}^{m}\left( k \beta{}^{2} \right)^{n}F_{n}^{\mathrm{(m)}}(\bm{q}).
    \label{eq:FmSF}
\end{equation}
The corresponding scattering intensity is calculated in accordance with  Eq.~(\ref{eq:intv2})
\begin{equation}
I_{m}^{\mathrm{(s)}}(q)/I_{m}^{\mathrm{(s)}}(0) = \left\langle | F_{m}^{\mathrm{(s)}}(\bm{q}) |^{2} \right\rangle,
    \label{eq:Imsf}
\end{equation}
where $I_{m}^{\mathrm{(s)}}(0) = n |\Delta \rho|^{2}S_{m}^{2}$, and $S_{m}$ is given by Eq.~\eqref{eq:areaSF}. The equation (\ref{eq:Imsf}) takes into account the spatial correlations between all the scattering disks composing the surface fractal.

In order to understand qualitatively the resulting scatting intensity (\ref{eq:Imsf}), we consider various approximations. When we neglect the correlations between \emph{mass fractal amplitudes} at various iterations, that is, $\left\langle F_{k}^{\mathrm{(m)}*}(\bm{q})F_{j}^{\mathrm{(m)}}(\bm{q}) \right\rangle \simeq 0$ for $k \neq j$, Eq.~\eqref{eq:Imsf} contains only the diagonal terms
\begin{equation}
\frac{I_{m}^{\mathrm{(s)}}(q)}{I_{m}^{\mathrm{(s)}}(0)}  \simeq \frac{\left( 1 - k \beta{}^{2} \right)^{2}}{\left( 1 - (k \beta{}^{2})^{m+1} \right)^{2}} \sum_{n=0}^{m}\left( k \beta{}^{2} \right)^{2n} \left\langle |F_{n}^{\mathrm{(m)}}(\bm{q})|^2 \right\rangle.
    \label{eq:Imsfv2}
\end{equation}
The approximation (\ref{eq:Imsfv2}) takes into account the correlations between the disk amplitudes \emph{within each mass-fractal iteration}, and it is sufficient to explain the exponent $d-D_\mathrm{s}$ in the SAS intensity of surface fractal (see the detailed analysis in Ref.~\cite{Cherny2017ScatteringFractals}).

One can move on and neglect the correlations between \emph{all the amplitudes of the disks}, composing GSCF. This approximation can be called the approximation of independent units. Then Eq.~\eqref{eq:Imsf} becomes the \emph{incoherent sum} of scattering intensities
\begin{equation}
\frac{I_m^{\mathrm{(s)}}(q)}{n |\Delta \rho|^{2}}\simeq \sum_{n=0}^{m} \beta{}^{n(4-D_\mathrm{s})}I_{0}(\beta^n_{\mathrm{s}}q).
\label{eq:Imsfv3}
\end{equation}
where $I_{0}(q)=S_{0}^2F_0^2(q)$ is the scattering intensity of disk of radius $r_0$. Here $F_0(q)$ is the formfactor of disk (\ref{eq:ffdisk}).

The approximation \eqref{eq:Imsfv3} helps us to easily understand the fractal power-law behaviour of the scattering intensity \cite{Cherny2017ScatteringFractals}. The intensity of disk $I_0(q)$ obeys the Porod law, i.e., $I_0(q)\simeq I_0(0)$ when $q\lesssim \pi/r_0$  and it decreases as $1/q^3$ when $q\gtrsim \pi/r_0$.  Since $\beta{}^{4-D_\mathrm{s}} \ll 1$, the first term dominates for $q\lesssim \pi/r_0$. However, at the point $q\simeq \pi/(\beta{} r_0)$ its contribution becomes about $1/\beta{}^{3}$ times smaller due to the $1/q^3$ decay, while the second terms is still remains the same. Thus the second term dominates at this point if the surface dimension obeys the inequality $4-D_\mathrm{s}<3$. Using the same arguments, we arrive at the conclusion that the $n$th term in Eq.~(\ref{eq:Imsfv3}) dominates at the point $q\simeq \pi/(\beta{}^{n-1}r_0)$. Therefore, increasing $q$ by $1/\beta{}$ times leads to decreasing the intensity by $1/\beta{}^{4-D_{\mathrm{s}}}$ times, and the slope of the scattering intensity on double logarithm scale is $\tau \equiv \ln \left( 1/\beta{}^{4-D_{\mathrm{s}}} \right) / \ln \left( 1/\beta{} \right) = 4-D_{\mathrm{s}}$. We arrive at the power-law behaviour (\ref{eq:int}), (\ref{eq:tau}) of surface fractal.

Note that for a couple of bound and randomly oriented point-like objects separated by the distance $l$, their amplitudes are correlated only for $q\lesssim 2 \pi / l$, while their correlator decays almost completely for lager $q$. In practice, one can explain any SAS intensity in terms of transitions from coherent to incoherent scattering\footnote{We use the term incoherent to describe various regimes of \emph{elastic}
scattering by analogy with optics, but not in the sense of ``the SANS cross section for incoherent scattering'', which gives a $q$-independent
background of SAS intensity (see, e.g., Ref.~\cite{book:bacon62}).} due to the decay of correlations between various objects, composing the sample. This analogy with optics is very useful for qualitative understanding of SAS curves \cite{Cherny2011DeterministicData}.

In particular, the analysis after Eq.~(\ref{eq:Imsfv3}) show that only the distributions of sizes of objects, composing a surface fractal, determine the scattering exponent $D_{\mathrm{s}}$, but the correlations between their amplitudes is not important here. The distribution of sizes of particles constituting a fractal (``internal polydispersity'') obeys the power-law (for random surface fractals) $dN(r)/dr \sim 1/r^\tau $, where $\tau = D_{\mathrm{s}}+ 1$  or its ``discrete'' analogy (for deterministic surface fractals) \cite{Cherny2017ScatteringFractals}. This suggests an answer to the long-standing question of whether it is possible to explain the scattering intensity for a surface fractal in terms of polydispersity \cite{Schmidt1982InterpretationVector}. The answer is ``Yes, it is possible approximately, if we are not interested in the `fine structure' of the scattering intensity''~\cite{Cherny2017ScatteringFractals,Cherny2017Small-angleSnowflake}.

\begin{figure}
	\begin{center}
		\includegraphics[width=\columnwidth,trim={0 0 0 0},clip]{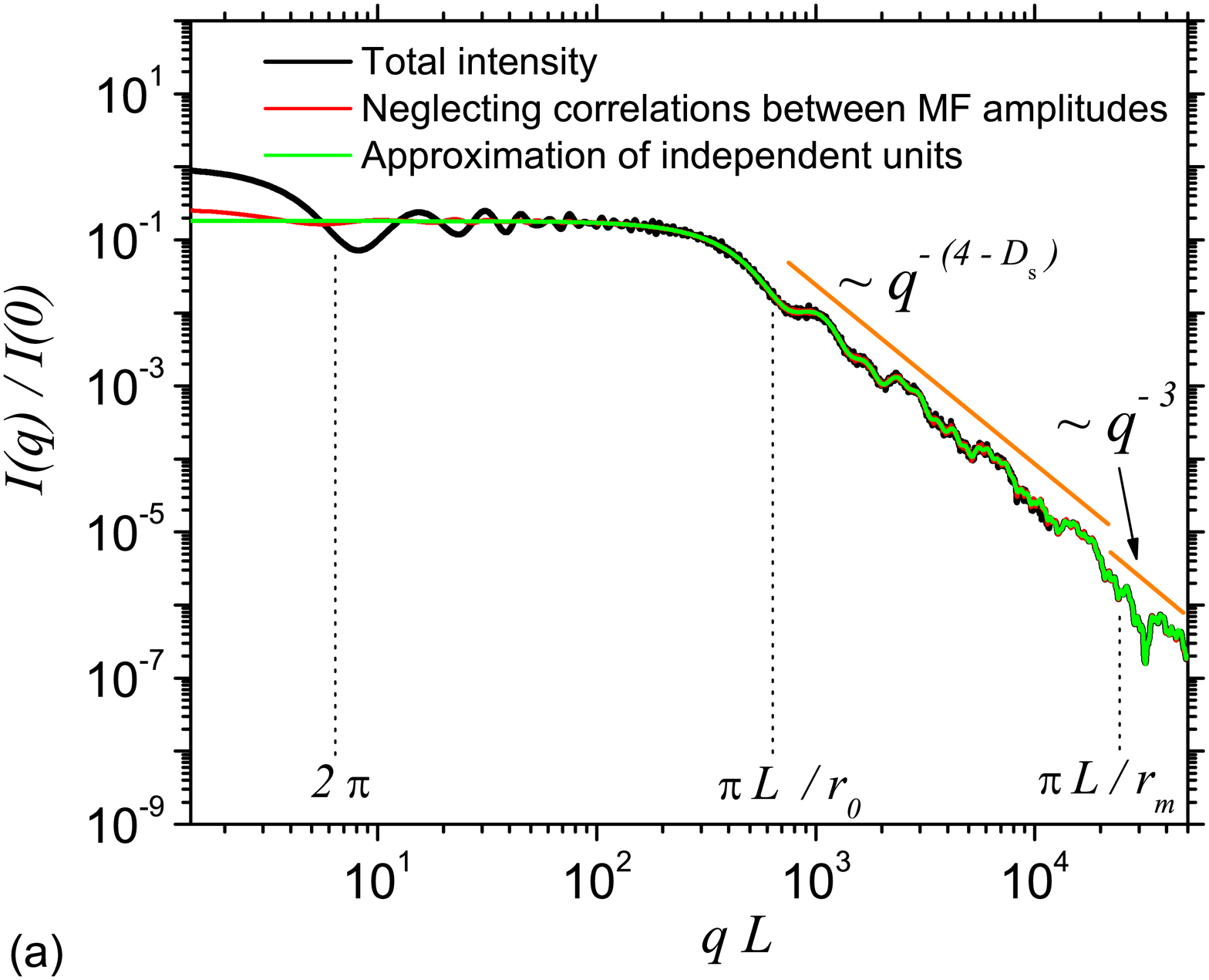}
		\includegraphics[width=\columnwidth,trim={0 0 0 0},clip]{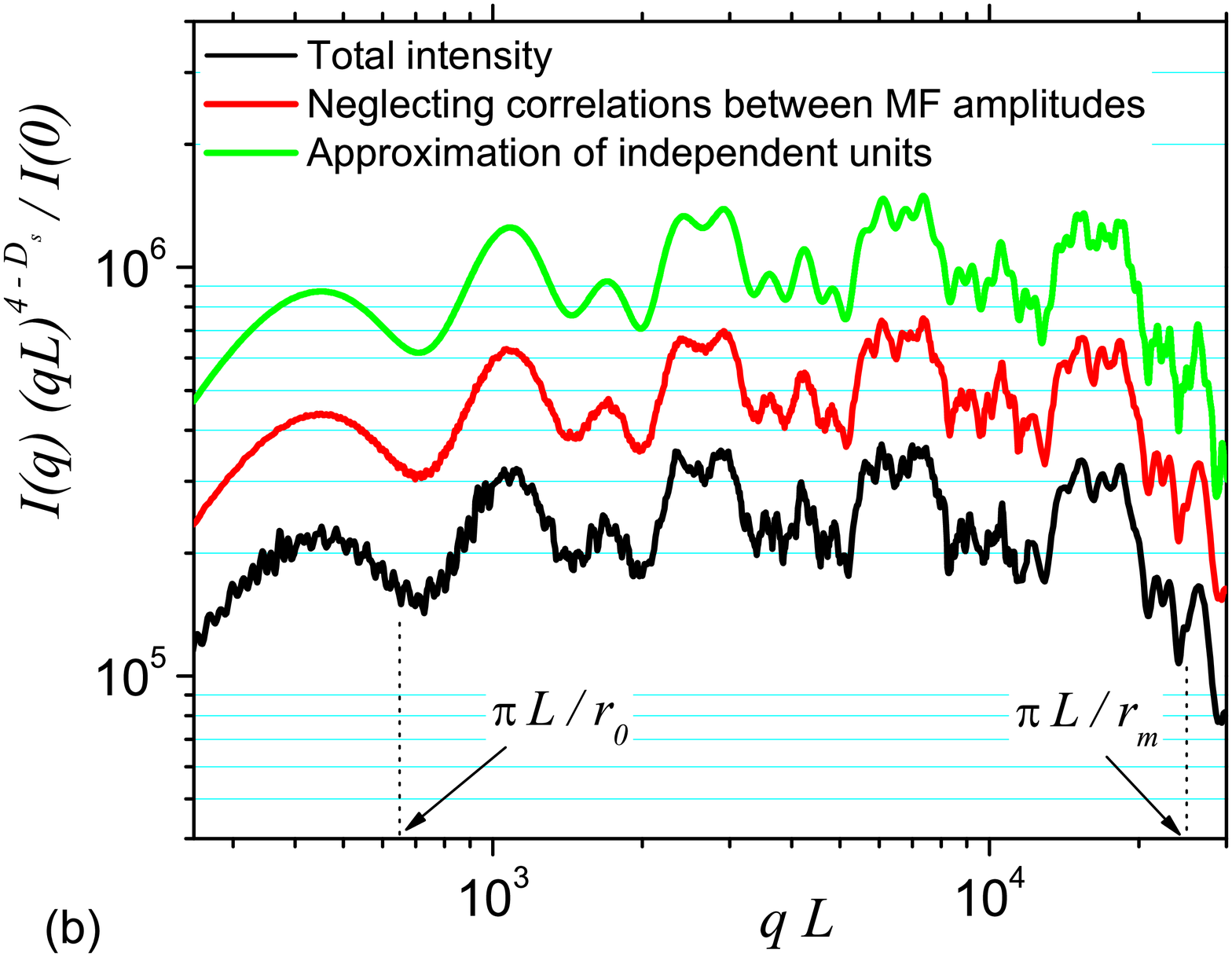}
		\caption{\label{fig:fig6} Monodisperse scattering from GSCF at $m = 4$ and $\beta = 0.4$ ($D_{\mathrm{s}} = 1.51\ldots$) for $h / l = 150$. (a) The black line shows the total scattering intensity given by Eqs.~(\ref{eq:FmSF}) and \eqref{eq:Imsf}, the red one represents its approximation \eqref{eq:Imsfv2}, which neglects the correlations between GMCF amplitudes, and the green line is the approximation of independent units \eqref{eq:Imsfv3}. (b) Approximate log-periodicity of the curve $I(q) q^{4-D_{\mathrm{s}}}$ in the fractal region. The red and green lines are scaled up for clarity by the factors 2 and 4, respectively.
}
\end{center}
\end{figure}

Figure \ref{fig:fig6} shows the total scattering intensity of GSCF \eqref{eq:Imsf}, the non-coherent sum of intensities \eqref{eq:Imsfv2} and the intensity when all the correlations are neglected \eqref{eq:Imsfv3} at scaling factor $\beta = 0.4$ ($D_{\mathrm{s}} \simeq 1.51\ldots$) for the fourth iteration. One can see the presence of four main structural regions. At $q \lesssim 2 \pi/L$ we have the Guinier region, as in the case of GMCF. At $2 \pi/L \lesssim q \lesssim 2 \pi / r_0$ we have a second shelf whose value coincides with  $I_{0}(0)$, which is the disk intensity at $q=0$. At $2 \pi / r_0 \lesssim q \lesssim 2 \pi / l$  (with $l = 2\beta{}^{4}r_0$), the fractal region occurs, where $I^{\mathrm{(s)}}(q) \sim q^{-(4-D_{\mathrm{S}})}$. Finally, at $q \gtrsim 2 \pi / l$, the Porod region is attained, as in the case of scattering from GMCF.

At $h / l = 150$, the total intensity is well approximated by the both approximations in the fractal region. This confirms the general observation \cite{Cherny2017ScatteringFractals} that the both approximations \eqref{eq:Imsfv2} and \eqref{eq:Imsfv3} are getting  better with increasing the ratio $h / l$. As seen, in the fractal region the effects of correlations between the spatial positions of the disks play a role only in additional oscillations, while the value of the scattering exponent is preserved. However, experimentally, these oscillations would be smeared out due to polydispersity effects or an instrumental resolution.

\subsection{Koch snowflake}

It may happen that the approximation of incoherent mass fractal amplitude (\ref{eq:Imsfv2}) does not work well. Since this approximation is crucial for explaining the exponent $d-D_{\mathrm{s}}$, it needs to be improved. We will show how to do this with a specific example.

\begin{figure}
\centerline{\includegraphics[width=0.7\columnwidth]{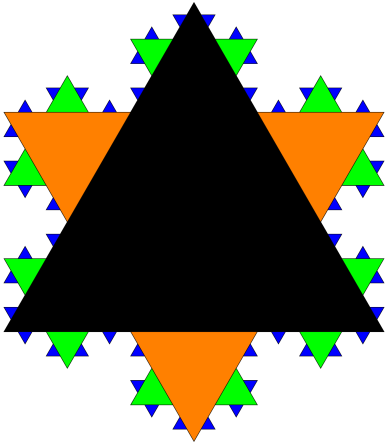}}
\caption{Koch snowflake as a sum of various iterations of mass fractals shown in different colors. Reproduced
from Ref.~\cite{Cherny2017Small-angleSnowflake}, Copyright \copyright{} 2017 The Royal Society of Chemistry.
}
\label{fig:fig3koch}
\end{figure}

Let us consider the well-known surface fractal called Koch snowflake (KS) following our paper \cite{Cherny2017Small-angleSnowflake}. In accordance with the main conjecture, it can be represented as a sum of mass-fractal iterations of the same dimension (see Fig.~\ref{fig:fig3koch}). The $n$th mass fractal iteration consists of triangles of equal sizes with the edge $a_n=a/3^n$, and their number is equal to $N_n=3\cdot 4^{n-1}$ for $n=1,2,\ldots$. Here $a$ is the edge of the largest (black) triangle in Fig.~\ref{fig:fig3koch}. The dimensions of the \emph{mass fractal} $D_{\mathrm{m}}$ and the \emph{perimeter} $D_{\mathrm{s}}$ coincide and are equal to $\lim_{n\to\infty}\ln N_{n}/\ln (a/a_{n}) = \ln4/\ln 3 = 1.26\ldots$. The area of the $n$th mass fractal iteration is given by $S_n=N_n \sqrt{3}a^2/(4\cdot3^{2n})$.

By construction, the KS amplitude can be written as the sum of the mass fractal amplitudes
\begin{equation}\label{KS_ampl_mass}
A_m(\bm{q})=\sum_{n=0}^{m+1} M_{n}(\bm{q}).
\end{equation}
The amplitudes are defined as $M_{n}(\bm{q})=\int_{S_n}\mathrm{d}\bm{r}\exp(-i \bm{q}\cdot\bm{r})$ and thus normalized at $q=0$ to the corresponding area: $M_{n}(0)=S_n$. The explicit analytical expressions for the KS amplitudes $A_m(\bm{q})$ can be found in Ref.~\cite{Cherny2017Small-angleSnowflake}.

\begin{figure}
\centerline{\includegraphics[width=0.85\columnwidth]{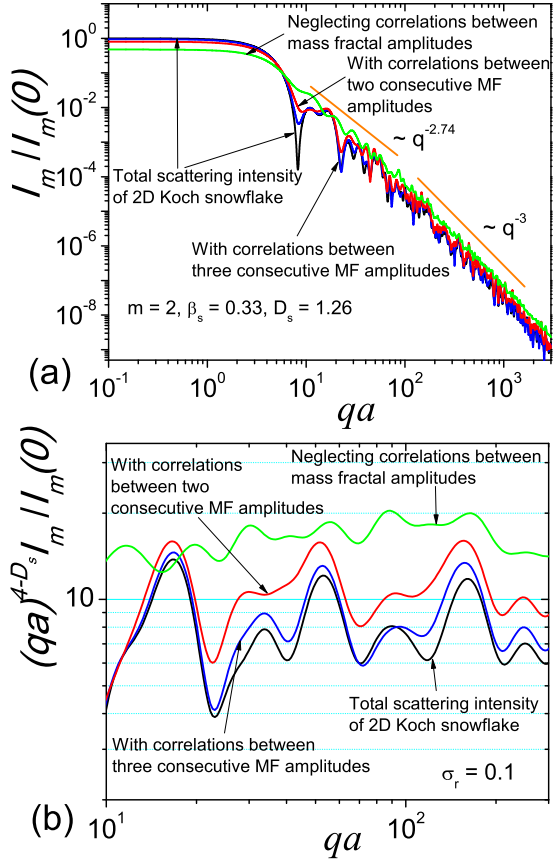}}
\caption{SAS from KS and its approximations by the scattering from the mass fractal composing KS (see Fig.~\ref{fig:fig3koch}). (a) The total intensity (black) and the intensities taking into account various correlations between the mass fractal amplitudes. The correlations between \emph{three} consecutive mass fractal amplitudes are included (blue), and the same for the correlations between \emph{two} consecutive mass fractal amplitudes (red). Neglecting all the correlations between mass fractals (green) is not good enough for describing the total intensity of KS. (b) Approximate log-periodicity of the curve $I(q)q^{4-D_{\mathrm{s}}}$  with the period $\beta=1/3$ (the power-law dependence is scaled out with $q^{4-D_{\mathrm{s}}}$). The polydisperse scattering intensities are shown for the relative variance $\sigma_{\mathrm{r}} = 0.1$. One can observe an interference between different mass fractal amplitudes, so their correlations are important. Reproduced
from Ref.~\cite{Cherny2017Small-angleSnowflake}, Copyright \copyright{} 2017 The Royal Society of Chemistry.}
\label{fig:fig6koch}
\end{figure}

The KS intensity\footnote{In this section, we omit the prefactor $n |\Delta \rho|^{2}$ in the definition of the SAS intensity (\ref{eq:intv2}).} $I_m(q)=\langle|A_m(\bm{q})|^2\rangle$ contains not only the mass fractal intensities $\langle|M_n(\bm{q})|^2\rangle$ but the correlations between the mass fractal amplitudes
\begin{align}
I_m&(q)=\sum_{n=0}^{m+1} \langle|M_{n}(\bm{q})|^2\rangle \nonumber\\
&+\sum_{0\leqslant n<p\leqslant m+1} \langle M^*_{n}(\bm{q})M_{p}(\bm{q})+M_{n}(\bm{q})M^*_{p}(\bm{q})\rangle.
\label{int_ampl_mass}
\end{align}
One can neglect the non-diagonal (interference) terms in this equation and even more, completely neglect the interference between the amplitudes of triangles composing the mass fractals. These approximations often work well, say, for the Cantor surface fractal (see Sec.~\ref{subsec:monoformfactor} above). However, Fig.~\ref{fig:fig6koch} shows that a complete ignoring correlations between the amplitudes of mass fractals leads to a poor approximation for the KS. The reason is that distances between  different mass fractal iterations and between triangles within one mass fractal iteration can be of order of their sizes, and we have to take into account the interference  terms in Eq.~(\ref{int_ampl_mass}).

Nevertheless, one can reduce the problem, in effect, to the incoherent sum of the ``combined" mass fractals.
Indeed, considering the correlations between two consecutive mass fractal iterations like $\langle M^*_{0}M_{1}\rangle$, $\langle M^*_{1}M_{2}\rangle$, and so on, and neglecting the other correlations, we obtain from Eq.~(\ref{int_ampl_mass})
\begin{align}
I_m(q)\simeq\sum_{n=0}^{m} \langle|M_{n}(\bm{q})+M_{n+1}(\bm{q})|^2\rangle -\sum_{n=1}^{m} \langle|M_{n}(\bm{q})|^2\rangle.
\label{int_double_mass}
\end{align}
SAS from the surface Cantor fractal is described well by incoherent sum of \emph{single} mass fractal intensities [see Eq.~(\ref{eq:Imsfv2})], while the first sum in the approximation (\ref{int_double_mass}) is nothing else but \emph{incoherent sum} of intensities of \emph{pairs} of consecutive amplitudes. The SAS intensities of each pair behave like a mass fractal with the power-law decay $I(q)\sim q^{-D_\mathrm{m}}$ at $D_\mathrm{m}=D_\mathrm{s}$, which results in the power-law decay of the intensity (\ref{int_double_mass}) $I(q)\sim q^{D_\mathrm{s}-2d}$ with $d=2$ for the plane.

By analogy with the \emph{pair} consecutive amplitudes, one can further improve the approximation (\ref{int_double_mass}) for the SAS intensity by including the \emph{triple} consecutive amplitudes $\langle|M_{n}+M_{n+1}+M_{n+2}|^2\rangle$. The results for the KS are shown in Fig.~\ref{fig:fig6koch}a.

One of the main properties of the SAS intensity is the approximate log-periodicity of the curve $I(q)q^{4-D_{\mathrm{s}}}$ within the fractal region, as shown in Fig.~\ref{fig:fig6koch}b.

\section{Generalizations}
\label{sec:fat}
\subsection{Mass fractals with heterogeneous scaling}
\label{sec:massGen}

The construction of GMCF in Sec.~\ref{sec:SASmass} suggests a generalization. Suppose that the iteration $m$ is composed of $k_m$ copies of the previous iteration $m-1$, scaled with the factor $\beta_{m}$. Then, following the same method, we find that its formfactor obeys the recurrence relation
\begin{align}\label{eq:Fn}
F_{m}(\bm{q})=G_m(\bm{q})F_{m-1}(\beta_{m}\bm{q}),
\end{align}
where
\begin{align}\label{eq:Gn}
G_{m}(\bm{q})=\frac{1}{k_m}\sum_{j=1}^{k_m}\exp(-i\bm{a}^{(m)}_{j}\cdot\bm{q})
\end{align}
with $\bm{a}^{(m)}_{j}$ being the positions of each copy of the iteration $m-1$. If we start from an initiator with the formfactor $F_0(\bm{q})$, the formfactor of the iteration $m$ can be written down explicitly with the help of Eq.~(\ref{eq:Fn})
\begin{align}\label{eq:Fnexpl}
F_{m}(\bm{q})=&G_{m}(\bm{q})G_{m-1}(\bm{q}\beta_{m}) G_{m-2}(\bm{q}\beta_{m}\beta_{m-1})\ldots  \nonumber\\
&\times G_{1}(\bm{q}\beta_{m}\beta_{m-1}\ldots\beta_{2}) F_0(\bm{q}\beta_{1}\ldots\beta_{m}).
\end{align}

This model can be realized in arbitrary Euclidean dimension $d$. For the $m$th iteration, the total number of primary objects is equal to $N_m=\prod_{j=1}^{m}k_{j}$, and their total volume in $d$-dimensions is given by $V_m=V_0\prod_{j=1}^{m}k_{j}\beta_{j}^{d}$, where $V_{0}$ is the volume of the initiator. The size of the objects composing the fractal is proportional to $\prod_{j=1}^{m}\beta_{j}$, and, hence, the fractal dimension is given by
\begin{equation}
D_{\mathrm{m}}= -\lim_{m \rightarrow \infty} \frac{\ln k_1+\ldots+\ln k_m}{\ln \beta_{1}+\ldots+\ln \beta_{m}}.
    \label{eq:DmGen}
\end{equation}
It follows from the Stolz--Ces\`aro theorem (see, e.g. Ref.~\cite{book:Muresan09}) that $D_{\mathrm{m}} = - \lim_{m\to\infty}\ln k_m/\ln \beta_{m}$ if the latter limit exists. The scattering intensity from an ensemble of the randomly distributed and oriented fractals is given by Eqs.~(\ref{eq:intv2}) and (\ref{eq:Fnexpl}). As usual, the structure factor is obtained in the particular case when the initiator is a point-like object with $F_{0}(\bm{q})=1$.

Let us discuss a few important features of the above construction. First, the total fractal can be composed of completely different fractals at each iteration, in general. For example, the first iteration can be the generalized Cantor fractal, the second the generalized Vicsec fractal, while the third the Menger sponge, etc. Thus, $G_{m}(\bm{q})$ are related to the positions of the structural units in \emph{the generator of the corresponding fractal}. Second, by construction, the iteration $m$ is composed of small copies of the iteration $m-1$, the iteration $m-1$ is composed of small copies of the iteration $m-2$, and so on up to the first iteration, which composed of the scaled copies of the initiator. Thus, the smallest composing ``blocks'' are related to the first iteration while the biggest ``blocks'' to the $m$th iteration.

This model is very general and adaptable. In the particular case of homogeneous scaling of the same fractal, $\beta_{m}=\beta=\text{const}$,  $G_{m}(\bm{q})$ is independent of $m$, and we arrive at a simple mass fractal with a single scale $\beta$, like the GMCF considered in Sec.~\ref{sec:SASmass}. If the generator is fixed but the scaling factor changes its value at some iteration $n\gg1$, we have a ``knee'' structure of the SAS intensity with different slopes in the fractal region on double logarithmic scale, which are associated with two different fractal dimensions. That means that the entire fractal is composed of structural units of a fractal with another dimension, and the both dimensions are observable at different ranges of the scattering vectors $q$. This is because the internal structure of an object is not visible if its size is smaller than $2\pi/q$. The detailed discussions of the ``knee'' structures in SAS curves and their origins can be found, e.g., in Ref.~\cite{cherny14JACr} (see also the next section below).

One can consider the set of the quantities $\beta_{m}$ and $k_{m}$ as fitting parameter to explain SAS curves and reveal the internal structure of various samples. This approach resembles a phenomenological model of Ref.~\cite{Beaucage1995ApproximationsScattering}.

\subsection{Fat fractals}

\begin{figure}
	\begin{center}
		\includegraphics[width=\columnwidth,trim={0 0 0 0},clip]{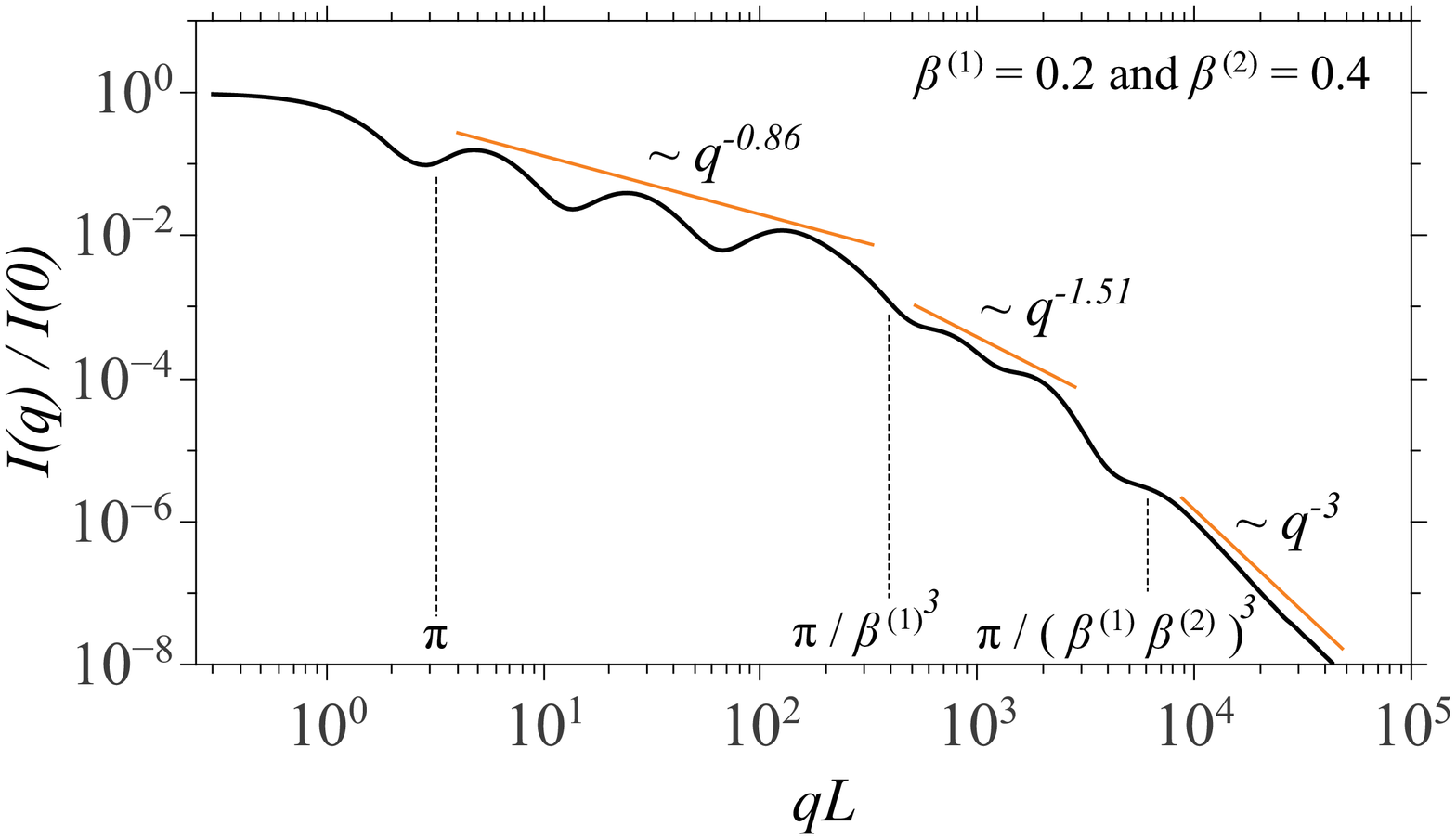}
		\caption{\label{fig:fig9} The scattering formfactors for the sixth iteration of the fat fractal in the presence of polydispersity (\ref{eq:DN})
with the relative variance $\sigma_{\mathrm{r}} = 0.3$. The diagram clearly reveals a succession of two power-law decays, whose scattering exponents $D_{\mathrm{m}}^{\mathrm{(1)}} = 0.86\ldots$ and $D_{\mathrm{m}}^{\mathrm{(2)}} = 1.51\ldots$
correspond to the ``fractal dimensions at each structural level.'' The vertical dotted-lines indicate the beginning and end of the two fractal regions, and the beginning of the Porod region.
}
\end{center}
\end{figure}

In this section, it is more convenient to use the inverse numbering of the generators in the fractal construction: ${1} \to {m}$, ${2} \to {m-1}$, $\ldots$,  ${m} \to {1}$. Then Eq.~(\ref{eq:Fnexpl}) reads
\begin{align}\label{eq:Fnexpl1}
F_{m}(\bm{q})=&G_{1}(\bm{q})G_{2}(\bm{q}\beta_{1}) G_{3}(\bm{q}\beta_{1}\beta_{2})\ldots  \nonumber\\
&\times G_{m}(\bm{q}\beta_{1}\beta_{2}\ldots\beta_{m-1}) F_0(\bm{q}\beta_{1}\ldots\beta_{m}).
\end{align}

The so called \emph{fat fractals} are a special case of mass fractals where the value of the scaling factor is allowed to increase with the iteration number $m$. This leads to sets with a fractal structure but non-zero standard measure (area, volume etc.). We illustrate the notion of fat fractals through a simple construction \cite{Anitas2014Small-angleFractals}, which is a particular case of the above general model of Sec.~\ref{sec:massGen}.

This construction is characterized by the sequence of scaling factors, which remain constant for three consecutive iterations and tend to one-half for $m\to\infty$
\begin{equation}
\beta_{m} = \frac{1 - c\,(1/3)^{p_{m}}}{2},
\label{eq:betasm}
\end{equation}
with $c$ being a positive constant smaller than one. The exponent $p_{m}$ is defined as $p_{m} = \left\lfloor{\frac{m-1}{3}}\right\rfloor$, where the function $\left\lfloor{x}\right\rfloor$ is the floor function (the greatest integer less than or equal to $x$). The fractal construction is the same as that of GMCF in Sec.~\ref{sec:SASmass} but with the scale $\beta_{m}$ at each iteration. Then the functions $G_{m}(\bm{q})$ in Eq.~(\ref{eq:Fnexpl1}) are given by
\begin{equation}
G_{m}(\bm{q}) = \cos \frac{q_{x}L(1-\beta_{m})}{2} \cos \frac{q_{y}L(1-\beta_{m})}{2}
    \label{eq:Gff}
\end{equation}
in two dimensions ($d=2$). It follows from Eqs.~(\ref{eq:DmGen}) and (\ref{eq:betasm}) with $k_{m}=4$ that the fractal dimension is equal to $D_{\mathrm{m}}=2$ in the limit of infinite iterations.

Figure \ref{fig:fig9}
shows the scattering from the sixth iteration of the fat fractal at the parameter $c=0.6$. It follows from Eq.~(\ref{eq:betasm}) that the scaling factor $0.2$ is kept constant at first three iterations, while it is equal to $0.4$ for the next three iterations. One can see the presence of two successive fractal regions, with the exponents $D_{\mathrm{m}}^{(1)} = 0.86\ldots$ and $D_{\mathrm{m}}^{(2)} = 1.51\ldots$, instead of a single fractal region, as in the case of scattering from GMCF\footnote{Here, the notation $(...)$ at the exponent is an index but not power.}. Thus the scaling factors can be interpreted as ``fractal dimensions at each structural level.''

The corresponding polydisperse scattering intensity (\ref{eq:intpoly})
clearly indicates that the values of the scattering exponents coincide with the expected values given by Eq.~\eqref{eq:Dm} with $\beta{}^{(1)} = 0.2$ and $\beta{}^{(2)} = 0.4$, respectively.
The oscillations of the intensity are smoothed when the polydispersity appears. Further increasing of the relative variance of the polydispersity $\sigma_{\mathrm{r}}$ leads to complete smoothing of the SAS curve.
As one can see, the scattering curve carries information about the fractal dimensions and the scaling factors at each structural levels.

\subsection{SAS from multifractals}

A single-scale deterministic fractal is composed of a basic geometric figure repeated on an ever reduced scale, and thus leading to a homogeneous structure. Inhomogeneous fractals considered in Sec.~\ref{sec:massGen} change the value of the scaling factor at each iteration, in general. However, fractals can have a heterogeneity of another type with rich scaling and self-similar properties, changing at every point,
known in the literature as multifractals.

\begin{figure}
	\begin{center}
		\includegraphics[width=\columnwidth]{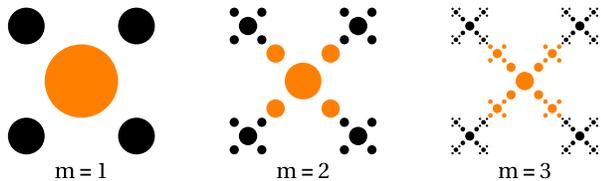}
		\caption{\label{fig:fig10} Iterations $m = 1, 2$ and 3 of the generalized Vicsek multi-fractal. Copies of antecedent iteration are scaled with the factors $\beta_{1}=0.25$ (black) and $\beta_{2}=0.5$ (orange).
}
\end{center}
\end{figure}

A good example of such fractals is a two-scale deterministic fractal, discussed in the Introduction. We give an explicit example similar to that of GMCF. Like GMCF, the initiator is a disk, and the first iteration includes four disks in the corners of the square at the same positions (\ref{eq:a}) with the scaling factor $\beta_{1}$. Besides, one more disk is added  in the center with another scaling factor $\beta_{2}$. At the second iteration, we repeat the same operation replacing the initiator by the first iteration, and so on. For $\beta_{1} = \beta_{2}$, we arrive at what is known in the literature as the Vicsek fractal, thus we name our model as a generalized Vicsek multi-fractal (GVMF). Figure \ref{fig:fig10} shows the first three iterations of GVMF with scaling factors $\beta_{1} = 0.25$ and $\beta_{2} = 0.5$ when the initiator is a disk.  In the limit of infinite number of iterations, we obtains the ideal GMVF, whose fractal dimension is given by Eq.~\eqref{eq:Mrv3} with $k_{1} = 4$ and $k_{2} = 1$, yielding $D = 1.36\ldots$.

The total volume in $d$-dimensions and the number of particles are equal to $(k_1 \beta_1^d + k_2\beta_2^d )^{m}V_{0}$ and $(k_1+k_2)^{m}$, respectively. Like in Sec.~\ref{sec:SASmass}, one can easily derive the recurrence relation between the subsequent iterations for arbitrary two-scale fractal in $d$ dimensions
\begin{equation}
F_{m}(\bm{q})\! =\! \frac{k_{1}\beta_{1}^{d} G_{1}(\bm{q})F_{m-1}(\beta_{1}\bm{q}) + k_{2}\beta_{2}^{d} G_{2}(\bm{q})F_{m-1}(\beta_{2}\bm{q})}{k_{1}\beta_{1}^{d}+k_{2}\beta_{2}^{d}}.
    \label{eq:GfGVMF}
\end{equation}
For the GVMF, $d=2$, $k_{1}=4$, $k_{2}=1$, the function $G_1(\bm{q})$ is given by Eq.~(\ref{eq:G1mf}) with $\beta=\beta_{1}$, and $G_2(\bm{q})=1$. With the relation (\ref{eq:GfGVMF}) and the initial formfactor (\ref{eq:ffdisk}), we obtain analytically the formfactor of arbitrary iteration.

The scattering intensity can be found in the standard way with Eqs.~(\ref{eq:GfGVMF}) and (\ref{eq:intv2}). For the point-like initiator whose density is described by the $\delta$-function, the SAS intensity can be found by averaging over the orientations $\langle|F_{m}(\bm{q})|^2\rangle$ with $F_{0}(\bm{q})=1$. After little algebra, we obtain its high-momentum asymptotics ($q \gg \frac{2\pi}{\beta_1^{m}L},\frac{2\pi}{\beta_2^{m}L}$)
\begin{equation}
\frac{I_{m}(q)}{I_{m}(0)}\simeq\frac{(k_1 \beta_1^{2d} + k_2\beta_2^{2d} )^m}{(k_1 \beta_1^d + k_2\beta_2^d )^{2m}} \label{eq:SasGVMF}
\end{equation}
for the $m$-th iteration.

\begin{figure}
	\begin{center}
		\includegraphics[width=\columnwidth]{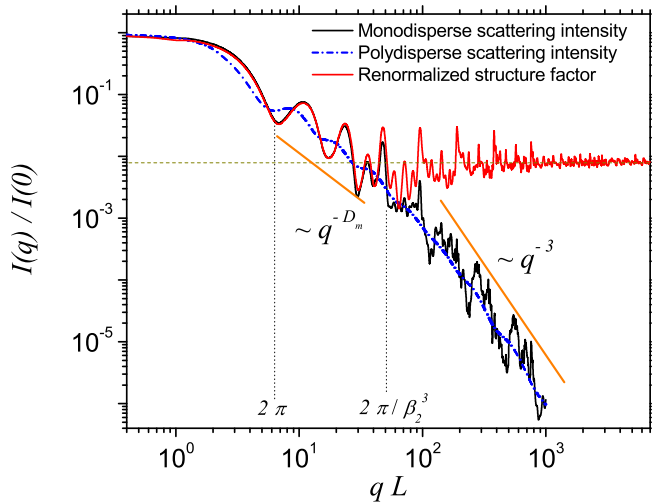}
		\caption{\label{fig:fig11} Monodisperse and polydisperse ($\sigma_{\mathrm{r}} = 0.15$) scattering intensities
from the two-scale multifractal model shown in Fig.~\ref{fig:fig10}. The asymptotics of the renormalized structure factor coincides with the theoretical value given by Eq.~\eqref{eq:SasGVMF}.
}
\end{center}
\end{figure}

Figure \ref{fig:fig11} shows the corresponding scattering intensity from GVMF at $m = 3$, shown in Fig.~\ref{fig:fig10}. The three main structural regions, i.e., the Guinier, fractal, and Porod ones are present, as expected. However, in the fractal region the periodicity of minima is not so obvious as in the case of scattering from GMCF, GSCF or fat fractals. Certainly, the reason of missing  any clear periodicity in the fractal region is the presence of two scaling factors at every scale. In this case, it is hardly possible to visualize the scaling factor directly from the SAS intensity. In principle, a multifractal can consist of a large number of scaling factors, and determination of all the scaling factors becomes an even more complicated task. Anyway, one can use the scattering formfactor for fitting the experimental data with the adjustable scaling factors and iteration number. At the same time, the scattering exponent in the fractal region coincides with the fractal dimension, as seen from Fig.~\ref{fig:fig11}.

\section{Future research}
\label{sec:future}

Although the description of SAS from deterministic fractals in recent years had considerable theoretical success, it is still not widely used in experiments. To be an effective approach to structural studies of deterministic fractals,
three main limitations should be overcome: materials, methods and instrumentation.

First, the widespread application of SAS to deterministic fractals is severely limited by the availability of the suitable samples. Currently, only several materials such as poly-silicone~\cite{Cerofolini2008FractalNanotechnology}, bis-terpyridine macromolecules~\cite{Newkome2006NanoassemblyGasketquot}, alkyletene dimers~\cite{Mayama2006MengerMethod}, single crystalline silicone~\cite{Berenschot2013FabricationSilicon}, Fe atoms, 1,3 -bis (4 - pyridyl) benzene molecules~\cite{Li2017ConstructionOrder}, terphenyl molecules~\cite{Li2015Sierpinski-triangleGroup} or dicarbonitrile molecules~\cite{Zhang2018SierpinskiAu111} have been used to create deterministic structures. However, they can not fulfill the ever increasing requirements for modern technological and industrial applications, such as  in stretchable electronics, nano- and micro-antenna, electrophysiological sensors, or precision monitors and actuators, and thus the types of deterministic materials need to be increased. In addition, the number of deterministic structures created from these chemical materials is quite low due to technological limitations or the high cost of raw materials, while a typical SAS experiment requires at least 10$^6$
deterministic fractals to be fabricated for each investigated sample. From a practical point of view, production of such a large number of fractals is still a challenge in most cases. Synthesis of the matrix in which the fractal is embedded, and the discovery of suitable mixing compositions between matrix and fractals are critical to increase the performances of materials with pre-defined structures and functions. Naturally occurring materials, such as heavy-water snowflakes are promising candidates to reduce materials costs and preparation time.

Although the developed deterministic models offer some specific methods for extracting additional information from SAS data, they actually imply knowledge of main features of their structure. This is a general drawback of SAS method, which actually gives the absolute value of the Fourier transform of the density, and thus the information about the phases is lost. Moreover, for randomly oriented objects, the squared Fourier transform is averaged over the orientations of $\bm{q}$. In this case, to retrieve the density from the output data is an ill-posed problem. However, measuring of the scattering from an ensemble of \emph{aligned} objects from \emph{different angles} with a \emph{position sensitive detector} makes this problem more viable. This approach is very similar to a computational imaging technique, known as Fourier ptychography (see, e.g., \cite{Zheng13}). The development of methods and approaches to this scheme is a difficult task both theoretically and experimentally.

Last, but not least, essentially all experimentally obtained curves are smeared out to a certain degree due to instrumental limitations that arise from wavelength spread, limits of the collimation or detector resolution. For deterministic fractals, this leads to reduced visibility of the oscillations in the fractal region and, as a consequence, the impossibility to extract several parameters of the fractals such as the scaling factor or the iteration number. Additional progress in SAS measuring instruments could help us eliminate the above-mentioned disadvantages to a greater degree, without increasing the measurement time and without sacrificing the complex morphology of the material under study.

\section{Conclusions}
For deterministic fractals, the SAS method was developed rapidly in recent years, both from the point of view of creating deterministic fractal structures at nano and micro levels, and in theoretical developments for analyzing the corresponding SAS intensity curves.
A large number of artificially created fractal structures at nano and microlevel motivates scientists to search for new applications in various fields of science, technology, industry and biomedicine. In this mini-review, we give some guidelines for a better understanding of the structural properties of deterministic fractals with a view to their further implementation in various applications. Further advances in the preparation of deterministic fractal materials, the development of their models, the analysis of SAS data, and the development of SAS tools can open new prospects for structural studies of SAS deterministic fractal structures.

\section{Acknowledgement}

The authors acknowledge support from the JINR--IFIN-HH projects.

\bibliography{rev}

\end{document}